\documentclass[a4paper,12pt,reqno,superscriptaddress,nofootinbib]{revtex4}
\usepackage[centertags]{amsmath}
\usepackage{amsfonts}
\usepackage{amssymb}
\usepackage{amsthm}
\usepackage{newlfont}
\usepackage{stmaryrd}
\usepackage{mathrsfs}
\usepackage{mathtools}
\usepackage{euscript}
\usepackage{graphicx}
\usepackage{enumerate}
\usepackage{todonotes}
\usepackage[normalem]{ulem} 

\usepackage{color}
\usepackage{floatrow}
\usepackage{caption}

\usepackage{tikz}
\usepackage{pgf}
\usetikzlibrary{positioning,fit,calc}
\usetikzlibrary{arrows,automata}
\usepackage{wrapfig}
\usepackage{subfigure}
\usepackage{amscd}
\usepackage{hyperref}

\theoremstyle{plain}
  \newtheorem{theorem}{Theorem}[section]

\theoremstyle{definition}

\theoremstyle{remark}
  \newtheorem{remark}[theorem]{Remark}

  \newtheorem{example}[theorem]{Example}

 \let\be=\beta  \let\ep=\epsilon
\let\ve=\varepsilon  \let\ga=\gamma 
\let\ka=\kappa

\newcommand{\caL}{{\mathcal L}}

\newcommand{\bbL}{{\mathbb L}}

\newcommand{\bbZ}{{\mathbb Z}}

\newcommand{\opunit}{\text{1}\kern-0.22em\text{l}}

\newcommand{\frj}{{\mathfrak j}}

\DeclareMathAlphabet{\mathpzc}{OT1}{pzc}{m}{it}

\newcommand{\id}{\textrm{d}}

\def\R{\mathbb R}

\let\oldsqrt\sqrt
\def\sqrt{\mathpalette\DHLhksqrt}
\def\DHLhksqrt#1#2{%
\setbox0=\hbox{$#1\oldsqrt{#2\,}$}\dimen0=\ht0
\advance\dimen0-0.2\ht0
\setbox2=\hbox{\vrule height\ht0 depth -\dimen0}%
{\box0\lower0.4pt\box2}}

\begin{document}

\title{Frenesy: time-symmetric dynamical activity in nonequilibria}

\author{Christian Maes\\{\it Instituut voor Theoretische Fysica, KU Leuven}}

\begin{abstract}
	We review the concept of dynamical ensembles in nonequilibrium statistical mechanics as specified from an action functional or Lagrangian on spacetime.  There, under local detailed balance, the breaking of time-reversal invariance is quantified via the entropy flux, and we revisit some of the consequences for fluctuation and response theory.
Frenesy is the time-symmetric part of the path-space action with respect to a reference process.  It collects the variable quiescence and dynamical activity as function of the system's trajectory, and as has been introduced under different forms in studies of nonequilibria.  We discuss its various realizations for physically inspired Markov jump and diffusion processes and why it matters a good deal  for nonequilibrium physics. This review then serves also as an introduction to the exploration of frenetic contributions in nonequilibrium phenomena.
\end{abstract}
\maketitle
\vspace{2cm}

\tableofcontents

\newpage
\section{Introduction}
Dissipative phenomena are central to nonequilibrium physics, be it in relaxational processes or in structure formation, in steady transport or in active matter. Entropy production rates are strictly positive there, \cite{dGM}, reflecting the Clausius heat theorem.  
Typical examples include heat conduction networks where a system is coupled to various thermal baths at different (constant) temperature, shear flow of a viscous liquid between plates moving at different speeds, molecular motors driven by the hydrolysis of an abundant ATP-concentration, or the inelastic bouncing of a ball heating up internal vibrational degrees of freedom.  There appear energy, particle or velocity currents in the system from which it is clear that time-reversibility is broken on the corresponding level of description.  It may also be that the currents flow in a chemo-mechanical configuration space and that the nonequilibrium gets visible in real space from the fluctuation behavior only.  For the latter we may have in mind membrane or flagella fluctuations in active matter (see e.g. \cite{rol}), or think about diffusion properties that are drastically changed by nonequilibrium effects (see e.g. \cite{bor,mud}).  More generally, in stationary nonequilibria  time-reversal symmetry gets broken on some level.  It has indeed been understood and since about two decades how precisely the physical entropy production for systems weakly coupled to equilibrium reservoirs is also a measure of time-symmetry breaking; see \cite{poincare,time,gibbs,crooks} for some of the original references.  For example, the relative entropy per unit time between a steady process and its time-reversal then exactly equals the mean entropy production rate, \cite{gibbs,time,jmp2000}. \\
Paradoxically however and quite generally, time-symmetric parameters and time-symmetric observables start to play a new and important role when sufficiently away from close-to-equilibrium situations.  Those variables and parameters are not thermodynamic but kinetic and it may thus appear that the type and mere number of nonequilibrium features grows so wildly that constructing a powerful nonequilibrium statistical mechanics outside the linear regime near equilibrium is hard to conceive. We may well wonder whether there is an alternative to a case by case detailed analysis of the dynamics.  We believe there is.  At least for some restricted class of nonequilibrium situations there does appear the possibility of a more systematic and less anecdotical approach. For that purpose, the theory goes on spacetime thereby constructing dynamical ensembles.  The possible paths or trajectories $\omega$ of the system in a spacetime window are weighted there via an action functional or Lagrangian function of $\omega$.  Its time-symmetric part is called the frenesy and the present paper will give a detailed account of its role in the study of some nonequilibria. More loosely, we will call certain time-symmetric observables such as (unoriented) traffic, residence times and dynamical activity, or even time-symmetric parameters such as prefactors in a reaction rate formula {\it frenetic}. As the concept is rather new, by its nature this paper will treat general features only and the examples or illustrations will be pedagogical.  This takes the form of a complimentary view to the emphasis on entropy production that has pervaded nonequilibrium thinking since its conception.  The purpose is to explain the role of time-symmetric (in)activity which we call the frenetic contribution in a variety of nonequilibrium phenomena.  At the same time we promote the view of a Lagrangian statistical mechanics where dynamical ensembles are defined on spacetime trajectories.  Just as the Hamiltonian is central to equilibrium statistical mechanics, so will entropy and frenesy combine to characterize the appropriate dynamical ensemble.  Apart from the conceptual or theoretically useful points, it must also be stressed that nowaday experimental aspects of nonequilibrium physics have moved far beyond macroscopic calorimetry or other thermodynamic measurements. In the 21st century many optical techniques often combined with data selection have been added to manipulate and resolve configuration paths in unprecedented ways. Today dynamical features including frenesy are becoming much more pronounced thanks to the ability to scan non-thermodynamic information (micro-fluctuations, colloidal trajectories, jump events in multilevel systems), and the development of a theory on frenesy is therefore timely.\\

Frenesy picks up aspects of time-symmetric dynamical activity and quiescence (with inactivity also considered as a form of activity in what follows).  The heuristics already appears for a macroscopic closed and isolated mechanical system.  Let us indeed revisit the usual Boltzmannian set-up where we fix the energy, the volume, the particle number and possibly other conserved quantities as constraints.  There is a many-to-one map which assigns to every ``microscopic'' state $\Gamma$ a condition $M=M(\Gamma)$.  Such an $M$ is most often a density profile of mass, of momentum or of some form of energy. 
The entropy of $M$ is obtained  by ``counting'' (in terms of the Liouville volume of the phase space region) the number of $\Gamma$ for which $M(\Gamma)=M$. If that number or (phase space) volume is $W(M)$, the entropy is $S(M) = k_B\log W(M)$, following Boltzmann. More than only explaining the Second Law in terms of initial and final equilibria, Boltzmann was arguing that during the mechanical evolution $S(M_t)$ grows to reach a maximum, with $M_t=M(\Gamma_t)$ for $\Gamma_t$ the micro-state at time $t$.  Indeed, when the physical coarse-graining in terms of the conditions $M$ has been done appropriately (see \cite{hthms}), then trajectories $\Gamma_t$ may be expected to reach into ever larger phase space volumes $M_t=M(\Gamma_t)$ when started from a random state $\Gamma_0$ prepared in a condition $M_0=M(\Gamma_0)$ with relatively small volume $W(M_0)$.  The dissipation function {\it par excellence} is the rate of change of phase space volume for the sequence of macroscopic conditions $M_t$, the entropy production rate.   Since we may identify $S(M_\text{eq})$ with the Clausius thermodynamic entropy, where $M_\text{eq}$ corresponds to the overwhelmingly largest phase space volume given the constraints, we get a first heuristic understanding of the second law.  In summary, that $S(M_t)$ is non-decreasing follows from volume-considerations where log-volume is entropy, typically to move mechanically from smaller to larger volumes, to reach the largest entropy given the constraints, called equilibrium. \\
Clearly that picture sketched briefly above where trajectories are pushed into macroscopic conditions $M$ with ever larger volumes $W(M)$ needs corrections even before discussing quantitative aspects. For one, that the size of the phase space region $M$ matters, usually expressed in terms of thermodynamic state variables, is only half the story, especially when still a long way from equilibrium.  E.g.~relaxation times are not (only) thermodynamic. The trajectory may spend a long time in a condition $M$ and not find ways to reach other 'neighboring' conditions, for example because of kinetic constraints built from disorder or induced by many-body interactions. Localization at high energies may for example prevent dissipation. That is expressed in periods of quiescence or of periodic breathing where knowing the expected escape rates would quantify the condition much further. As a way of speaking, not only the size (volume) but also the exit/entrance area (surface) counts, especially when the trajectory moves in smaller nonequilibrium phase space regions. Moreover, passing from one condition to another also has a time-symmetric traffic component which is invisible in the change of volumes (or entropies).  Frenesy wants to quantify that ``lingering and trafficking.''   Surely far from equilibrium, in conditions with relatively small phase space volumes, we may expect that frenesy complements entropic considerations.  The frenetic contribution will then reach equal importance to the entropic one, thence its relevance for nonequilibrium physics.\\  Secondly, today we manage to monitor ``mesoscopic'' states $x$, which live on a much finer scale of coarse graining than for the usual thermodynamics. On that third level of description, between the microscopic laws and the macroscopic behavior, fluctuations can be made visible and hierarchies of time and length scales appear, \cite{3lev2}.  Without macroscopic limit there is e.g. no unique mesoscopic state $x$ representing equilibrium; rather there is an equilibrium statistical occupation over various $x$.  Fluctuations, hence also their time-symmetric sector, are essential on those scales.  Response theory must take into account those time-symmetric fluctuations and not only the fluctuations of and correlations with dissipation functions (as is the miraculous case for linear response around equilibrium).\\

The previous heuristics becomes more concrete and less abstract in the following sections.  Yet, the value of the concept of frenesy is in unifying a number of aspects and to reduce certain phenomena to their frenetic origin.    We start in the next section with a brief orientation about the meaning of nonequilibrium, if only to avoid small problems with terminology.  We continue in Section \ref{timesym} with the introduction of key time-symmetric parameters and observables. Those ``activity'' parameters are able to select occupation and current characteristics when out-of-equilibrium. Lifting the analyis to spacetime ensembles leads to the formal definition of frenesy in the context of Markov processes. By that very definition frenesy takes a central role in dynamical fluctuation theory as is explained in Section \ref{dynf}. E.g. excess escape rates allow the formulation of a variational problem for stationary nonequilibria.  Section \ref{res} discusses the frenetic contribution in response theory.  That includes linear response around nonequilibria as well as nonlinear response around equilibrium.  We also comment there on some particulars of statistical forces, i.e., the induced force on a probe in a nonequilibrium environment.  The Einstein relation between noise and friction is broken due to the frenetic term in linear response.  Two shorter sections end the review.   Section \ref{dynsmf} searches for the analogue of frenesy in smooth dynamical systems.  We have in mind the example of the standard thermodynamic formalism where e.g. entropy production rate is identified with phase space contraction rate.  Finally, in Section \ref{expf} we list a number of experimental studies where the measurability of frenesy has been proven.  Especially today with the exceptional possibilities of manipulating trajectories on micrometer scales there are interesting and important opportunities to observe frenesy quasi-directly.

\section{What is nonequilibrium?}
Nonequilibria are ubiquitous from the largest scales of cosmology, over astrophysics and plasma physics, to atmosphere dynamics, turbulence and rheology, to life phenomena and micro-biology all the way to quantum hydrodynamics and nanoscopic transport.  Briefly stated, we recognize nonequilibrium processes from the breaking of time-reversal symmetry.  There are currents in energy, particle, momentum or volume to name the most general cases. When maintained in a steady state, the system must be open to exchanges of these same quantities with the external world.  The system then maintains a nonequilibrium condition where again energy, particles,...  are dissipated in that environment.\\
To get some conceptual orientation it probably makes sense to distinguish three stages of nonequilibrium physics, somewhat in chronological order but overlapping in essential ways and each being realized today as active fields of research:\\
\begin{enumerate}
\item 
Nonequilibrium theory provides the basis for understanding the stability of thermodynamic equilibrium.  Initial studies in nonequilibrium physics certainly emphasized the problem of relaxation and return to equilibrium.   On mesoscopic levels of description we deal then with dynamics that satisfy the condition of detailed balance implementing the microscopic reversibility. The equilibrium condition is characterized dynamically as one of zero entropy production, while the entropy production rate is positive outside equilibrium.  Linear response theory around equilibrium expresses a fluctuation--dissipation relation. On a macroscopic level, the dynamics for returning to equilibrium are characterized by gradient and/or GENERIC flow \cite{ott,gen1}.  It is still a major program in nonequilibrium statistical mechanics to clarify mathematically the origin of macroscopic irreversible behavior.
Other fascinations include localization phenomena and glassy dynamics, which have important kinetic ingredients as well.
\item When a system is in contact with well-separated different equilibrium reservoirs to exchange energy, particles, volume or momentum, we may observe steady transport and dissipation.  Here we use a local version of detailed balance to characterize mesoscopic dynamics; see already in the next section.  Another option is to drive the system via some time-dependent protocol and when periodic, the system may again enter in a nonequilibrium steady state.  There is a possibility to quantify ``the distance from equilibrium'' via the amplitude of driving nonconservative forces, or via the amount of time-reversal breaking in the steady state (e.g. measured via relative entropy between the process and its time-reversal).  For the dynamics, the violation of (global) detailed balance is taken as the key-signature of nonequilibrium.  Steady state or nonequilibrium thermodynamics asks for a macroscopic characterization of resulting nonequilibrium phases \cite{sst}.   For small open systems nonequilibrium fluctuation theory is the central object of study.
\item A system (like a collection of probes) can also be in contact with a nonequilibrium medium of the type above.  Then, particles couple their motion to nonequilibrium degrees of freedom.  A possible example are active particles where spatial translation is coupled to chemo-mechanical and driven degrees of freedom.   Also local detailed balance gets violated now, and the nonequilibrium extends over a larger number of scales and levels of descriptions. Many of the problems of biophysics arise exactly there. At that moment, the notion of entropy production refers to dissipation in a more distant equilibrium reservoir, which is for example thermostatting a nonequilibrium medium to which the particles are coupled. Questions on statistical forces, fluctuation and response behavior and the derivation of Brownian motion induced by contact with nonequilibria are central here \cite{stefan}.
\end{enumerate}

For what follows in the next sections we will be mostly focusing on steady nonequilibria where the condition of local detailed balance is satisfied.  Local detailed balance is a condition of local dynamical equilibrium, in the sense that the considered updates of the system are either Hamiltonian or each time involve exchanges of energy or particles with one equilibrium reservoir at the time only.  One should always keep in mind here that words such as ``equilibrium,'' ``local'' and even ``system'' or ``coupling'' implicitly refer to a certain scale of description.  Equilibrium {\it versus} nonequilibrium is not a purely dynamical distinction but depends on levels of physical coarse-graining as well.  Similarly, whether a process breaks time-reversal symmetry is in the eye of the beholder.  That is not to say that such notions are subjective as we most often easily agree for good physical reasons on the appropriate levels of description and on the collection of relevant variables and manipulations.

\section{Time-symmetric kinetics}\label{timesym}
As alluded at above we are interested in a mesoscopic level of description.  The states $x_t$ may be positions or more general system configurations that evolve stochastically because many degrees of freedom have been integrated out (e.g. in some weak coupling limit) that have an initial (e.g. thermal) distribution.  Another possibility is that we are considering the approximation for quantum dynamics on the level of the Dirac-Fermi Golden Rule.  Discrete state spaces and indistinguishable particles are normal situations then.\\
For mathematical concreteness we consider here first the framework of Markov jump processes on a finite state space $K$ with elements (states) $x,y,\ldots $. In that way, for the moment, we pay no special attention to spatial extension or to interactions. A typical context is chemical kinetics or the evolution of spin degrees of freedom.  The $x$ may also denote energy levels or specific chemo-mechanical configurations of a molecule.  Throughout we write $k(x,y)$ for  transition rates in the jumps $x\rightarrow y$ and we assume that the process is irreducible. The unique stationary distribution is denoted by $\nu$, solution of the stationary Master equation, $\sum_{x,y} j_\nu(x,y) =0$, for the expected directed bond-current 
\begin{equation}\label{dbc}
j_\rho(x,y) := \rho(x) k(x,y) - \rho(y)k(y,x)
\end{equation} 
when the density is $\rho$.  The words `density' and `distribution' refer to the trivial thermodynamic description  for the statistical occupation of the states, imagining many independent random walkers $x_t$ on $K$.  While all that looks extremely simple we still believe it to be a good starting point, perhaps in the sense of the following quote by David Ruelle \cite{ext},
\begin{quote}
	My inclination is to postpone the study of the large--system
		limit: Since it is feasible to analyze the nonequilibrium
		properties of finite systems —--- as Gibbs did for their equilibrium
		properties --- it seems a good idea to start there.\\
		That may not answer all questions, but it advances nonequilibrium
		statistical mechanics to the point equilibrium
		had reached after Gibbs.
	\end{quote}

Let us give the transition rates for the jumps $x\rightarrow y$ via the parameterization
\begin{equation}\label{gfo}
k(x,y) = a(x,y)\,e^{s(x,y)/2}
\end{equation}
where the activity parameters are symmetric,
\begin{equation}
	\label{apa}
	a(x,y) = a(y,x) = \sqrt{k(x,y)k(y,x)}
\end{equation}
and the
\[
s(x,y) = -s(y,x) = \log \frac{k(x,y)}{k(y,x)}
\]
are antisymmetric.  Under the condition of local detailed balance \cite{go,hal,derrida,leb,time,snak}, one imagines an environment consisting of spatially separated equilibrium baths, each with fast relaxation.  Then, those $s(x,y)$ give the (discrete) change of entropy per $k_B$ in the equilibrium baths with which energy, volume or particles are exchanged during the system transition $x\rightarrow y$.   It even makes sense then to call
\begin{equation}
	\label{enp}
	S(\omega) = \sum_s s(x_{s^-},x_s)
\end{equation}
the path-wise entropy flux (per $k_B$) in the environment.  The sum is over all the jump times in the (system) trajectory $\omega = (x_s, 0\leq s\leq t)$ over the time $[0,t]$, with $x_{s^-}$ the state just before the jump to $x_s$ at time $s$.  Note that we thus read the variable changes of the entropy in the reservoirs in terms of system trajectories.  That is highly nontrivial and requires (e.g. weak coupling and separability) assumptions which is however not the topic of the present review.  See \cite{gibbs,jmp2000,time} for its introduction in steady nonequilibrium statistical physics.  In the transient regime of time-dependent detailed balance dynamics which is the context of the Jarzynski relation, there is a similar ``thermodynamic'' formula introduced by Crooks, \cite{crooks}.\\
Obviously, a lot of physics and suitable approximations are imagined here, and are mentally added to the mathematics.  The point is of course that those Markov processes are {\it derived} in some limiting regimes from a more microscopic dynamics after some physics inspired coarse-graining. \\

Note that in general both the $s(x,y)$ and the $a(x,y)$ are function of system parameters and external driving, temperature etc.  It is also realistic and natural to influence the $a(x,y)$ without touching the $s(x,y)$.   In biological systems, for a given source of low entropy as defined e.g. by the environment, an organism could change its kinetic parameters to select a certain condition or to adapt to challenges. It is indeed correct to think that the activity parameters $a(x,y)$ enter the stationary distribution $\nu$ whenever they vary over the bonds $(x,y)$ but surprisingly, that is not the case under (global) detailed balance.  As is well known indeed, whenever there is a potential $V$ so that for all bonds $(x,y)$ where $a(x,y)\neq 0$ we have $s(x,y) = \left(V(x) - V(y)\right)/(k_B T)$, then the stationary distribution is the Gibbs equilibrium $\nu_\text{eq} \sim \exp -V/(k_BT)$ for potential $V$ at temperature $T$. Properties of relaxation to that $\nu_\text{eq}$ do depend on the $a(x,y)$'s but the stationary occupation fluctuations do not.
The activities $a(x,y)$'s  do however become crucially relevant also for stationary properties from the moment the condition of (global) detailed balance gets broken. That concerns both occupation and current selections. Network control, either by humans, machines or by nature at large, will therefore be able to take advantage of managing time-symmetric kinetics for creating handles to selectivity. 

\subsection{Population selection}\label{pops}

Nonequilibrium aspects may reveal the specific kinetics that was already there under detailed balance but was unnoticed in the equilibrium distribution.  Specifically, the $a(x,y)$'s of above may have an important impact on population selection.  That is not surprising from a mathematical point of view, but it remains physically interesting to see how that works and how the occupations indeed may be selected via time-symmetric kinetics.  There is a wide plethora of examples ranging from cosmology to micro-biology. We can think about Fermi acceleration for the abundance of high-energy particles and the origin of cosmic radiation \cite{Fer} or about the mechanism of kinetic proofreading for assuring the appropriate relative expression of proteins or for obtaining increased specificity, \cite{hopfield}.\\ 
Here we present some elementary considerations and examples.

\begin{example} [blowtorch theorem]
The blowtorch theorem is not so much a mathematical theorem as it is a physical understanding that entropy production principles or versions of heterogeneous thermodynamics alone cannot suffice to identify steady-state distributions in the more interesting regimes of nonequilibrium physics.  The idea goes back to Landauer; see \cite{land1975}.   In the paper \cite{heatbounds}  the following more precise formulation is chosen.\\
Jump processes as above in \eqref{gfo} can be seen as random walks on a graph ${\cal G}$.  The vertices are the states $x\in K$ and the oriented edges are made by the bonds $(x,y)$ where $a(x,y)\neq 0$.     
The process takes oriented paths on ${\cal G}$ where the random walker jumps over a sequence of bonds $(x_0,x_1),(x_1,x_2),\ldots,(x_{n-2},x_{n-1}),(x_{n-1},x_n)$. For the Markov process we simply denote a path by $\omega$ like in \eqref{enp} and we write $S(\omega)$ for the total entropy flux (per $k_B$): $S(\omega) := s(x_0,x_1) + s(x_1,x_2) +\ldots + s(x_{n-2},x_{n-1}) + s(x_{n-1},x_n)$.  In the case of (global) detailed balance, 
\[
S(\omega) =  \frac 1{k_BT}\,\left(V(x_0) - V(x_n)\right), \qquad \text{ when  }  \;\; s(x,y) = \left(V(x) - V(y)\right)/(k_B T)
\]
 but otherwise it depends on the details of the path, and not only on its beginning and end.  We introduce next a {\it heat order}, writing  $x \succeq y$ whenever $S(\omega) \geq 0$ for all paths $\omega$ from $y=x_0$ to $x=x_n$.  I.e., two states are heat ordered when {\it all} ways of going from one to the other are dissipative (either all exo- or all endothermic). 
It is shown in \cite{heatbounds} that $x \succeq y$ implies $\nu(x) \geq \nu(y)$ in the stationary distribution $\nu$, even when there is no detailed balance.  Yet it is common in natural models to find many pairs of states $x^*$ and $y^*$ that are not heat-ordered:  there is a path $\omega_1: x^*\longrightarrow y^*$ with $S(\omega_1)\geq 0$, and there is a path $\omega_2: x^*\longrightarrow y^*$ with $S(\omega_2)< 0$.  Then neither $x^* \succeq y^*$ nor $y^* \succeq x^*$ holds.  The blowtorch theorem of \cite{heatbounds} states that in that case we can always produce either $\nu(x^*) > \nu(y^*)$  or $\nu(x^*) < \nu(y^*)$ by (only) changing the time-symmetric $a(x,y)$'s.\\
In other words, the time-symmetric activity parameters decide on the stationary relative occupation when the states are not heat-ordered.  That has interesting implications and provides possible applications.  We or nature indeed can start to manufacture the time-symmetric kinetics to produce the required, correct or best levels of population.  Some levels may become over-abundant with respect to equilibrium.  That occurs in many natural systems but also in more extreme cases of population inversion.  Here comes first the simplest set-up as described first in \cite{urna,winny}.\\
\end{example}

\begin{example}
	[population inversion]\label{inp}
Consider a multilevel system where energies $E(x)$ are associated to states $x = 1,2,\ldots, N$.  The detailed balance Markov jump process would have purely thermal transition rates,
\[
k(x+1,x) =  a(x+1,x) \,\exp (E(x) - E(x+1))/(2k_BT)
\]
for the jumps (only) between neighboring energy levels in an equilibrium bath at temperature $T$.  We think of independent particles occupying those different (non-degenerate) energy levels.  The constants $a(x,x+1)=a(x+1,x)$ parameterize the undirected activity between levels $x$ and $x+1$ and possibly still depend on energies and temperature, much like in Arrhenius to Kramers reaction rate theory, \cite{hang}.
The stationary occupation follows the Gibbs equilibrium prescription $\nu_\text{eq}(x) = \frac 1{Z} \exp -E(x)/k_BT$, independent of the kinetic (time-symmetric) activity parameters $a(x,y)$.  Let us for simplicity assume that the energies can be linearly ordered, $E(1) < E(2) \ldots < E(N)$.  At low temperature (compared to the energy differences) mostly the low lying energy levels (say states $x=1$ and $x=2$) will be occupied.\\
In contrast with that Boltzmann-Gibbs scenario of occupation we now want to overpopulate the higher energy levels.  For that purpose we drive the system by energizing particles.  We can do that by introducing a (new) transition between levels 1 and $N$ (the lowest and highest energy):
\[
k(1,N) = b = k(N,1) > 0
\]
thereby creating a loop in the graph of transitions.  Now the particles undergo both big losses and big gains of energy symmetrically, by an external mechanism (pump): the particles move back and forth between the lowest and the highest energy levels.  Of course, once they reach the highest energy they could cascade down thermally and fast at sufficiently low temperature $T$.   Yet, the system is out-of-equilibrium now with a new stationary distribution $\nu$ where the reactivities or activity parameters $a(x,x+1)$ start to matter much, especially at large $b$ and low $k_BT$. Let us indeed make thermal escape from the highest energy level difficult by choosing
\[
a(N,N-1) = e^{-F/k_BT}\, \exp (E(N-1) - E(N))/(2k_BT)
\]
with large barrier $F/k_BT \gg 1$. The particles get essentially trapped at the highest energy level, after being sent there at rate $b$ from the lowest energy level. There only remains an infinite temperature bridge between the highest and the lowest energy level.  Then, the new stationary distribution $\nu$ (depending also on $b$ and $F$) shows population inversion in the sense that the highest energy level will be highly occupied as $b$ and $F/(k_BT)$ grow large, leading to an effective temperature
\begin{equation}\label{tef}
T_\text{eff}(N) := \frac{E(N)-E(1)}{k_B}\,(\log \frac{\nu(1)}{\nu(N)})^{-1} \gg T
\end{equation}
as a measure of occupation of the high-energy level compared to the ground state.\\
\end{example}

Similar considerations can be made in a more general set-up where the escape rates gradually decrease as the energy gets larger. In fact, in
exactly that same way we get a general scenario for the origin of suprathermal tails in nonequilibrium velocity distributions $\nu(\vec v)$, \cite{nonMax}.\\  

\begin{example}
	[Nonmaxwellians]
We leave for a moment the context of jump processes and we consider the
Fokker-Planck equation for the time-dependent velocity distribution $\rho(\vec v,t)$ of a dilute gas in three spatial dimensions:
\begin{equation}\label{fknon}
\frac{\partial \rho}{\partial t}(\vec v,t) =
\frac 1{v^2}\frac{\partial}{\partial v}\left(\frac{\gamma(v)}{m}\,v^2\,[v\,\rho(\vec v,t) + \frac{k_BT}{m}\frac{\partial \rho}{\partial v}(\vec v,t)]\right) +   \frac 1{v^2}\frac{\partial}{\partial v}\left(v^2\,\frac{B(v)}{m^2}\frac{\partial \rho}{\partial v}(\vec v,t)\right) 
\end{equation}
When $B(v)=0$ there is detailed balance, and the Maxwell distribution $\nu_\text{Max}(\vec v)$ appears as the stationary solution of \eqref{fknon},
\[
v\,\nu_\text{Max} + \frac{k_BT}{m}\frac{\partial \nu_\text{Max}}{\partial v} = 0 
\]
independent of the choice of the friction parameter $\gamma(v) >0$.  Yet, from the moment we add the second diffusion in \eqref{fknon}, taking $B(v) \neq 0$, the system falls out of equilibrium and the specific dependence of $\gamma(v)$ on $v$ starts to matter greatly.  We may easily get localization at high energies, which results in suprathermal tails for the nonequilibrium stationary velocity distribution.  The $\gamma(v)$  and the $B(v)$ play here the role of kinetic time-symmetric parameters, similar to the $a(x,y)$ with $F$ and $b$ in the above jump processes.  As an example, if $\gamma(v) \propto v^{-3}, B(v) \propto v^{-1}$ for large $v$, then the stationary solution $\nu(\vec v)$ of \eqref{fknon} verifies a  power-law decay for large speeds:
\begin{equation}\label{resu}
\nu(\vec v) \propto \frac{1}{v^{2\kappa}},\qquad v\uparrow \infty
\end{equation}
with power-law exponent $\kappa$ being determined by further parameters such as mass, equilibrium temperature and the nature of the driving.
That is what really happens in space plasmas where those kappa-distributions \eqref{resu} have indeed been observed and are well established, \cite{peir}.  There, the $\gamma(v)$ is derived from Coulomb scattering and the $B(v)$ gets its dependence from the diffusive acceleration of electrons or ions in time-dependent electromagnetic fields in a turbulent plasma; see \cite{nonMax}. 
\end{example}

\subsection{Current characteristic}

Also various aspects of a steady current may be influenced by the time-symmetric $a(x,y)$'s. That can refer to the magnitude of the current of course, but also to its direction. We give two simple examples.  The first one is an active Browian particle moving in a landscape which is governed by an internal three-state process.  A similar toy-model is described in \cite{nong}.

\begin{example}\label{crex}[Generating emf]
Consider a probe moving diffusively on the ring, $Y \in S^1$, and coupled to a driven system with three states $x = 0, 1, 2$.  In that sense the probe is an active Brownian particle as its translation is coupled to another (discrete internal) degree of freedom.  We imagine the coupling via an interaction energy $U(Y,x)$ and for the three-state system we take the transition rates of the general form \eqref{gfo},
\begin{align}\label{rl}
k(0,1) &= a(Y)\,e^{\frac{\be}{2}\,[U(Y,0) - U(Y,1) + \ve]},\quad k(1,0) = a(Y)\,e^{\frac{\be}{2}\,[U(Y,1) - U(Y,0) - \ve]}\nonumber
\\
k(1,2) &= b(Y)\,e^{\frac{\be}{2}\,[U(Y,1) - U(Y,2) + \ve]},\quad k(2,1) = b(Y)\,e^{\frac{\be}{2}\,[U(Y,2) - U(Y,1) - \ve]}
\\
k(2,0) &= c(Y)\,e^{\frac{\be}{2}\,[U(Y,2) - U(Y,0) + \ve]},\quad k(0,2) = c(Y)\,e^{\frac{\be}{2}\,[U(Y,0) - U(Y,2) - \ve]}\nonumber
\end{align}
The parameter
$\ve\geq 0$ is the driving amplitude along the cycle
$0 \rightarrow 1 \rightarrow 2$.  Observe that the activity parameters $a(Y), b(Y), c(Y)$ are made to depend on the position $Y$ of the probe.  All functions of $Y$ 
are periodic obviously, and a thermal bath at inverse temperature $\beta\neq 0$ plays the role of environment.  The rates \eqref{rl} satisfy local detailed balance with $\ve\neq 0$ causing time-reversal breaking (in irreversible dissipation).  The stationary distribution of \eqref{rl} at fixed probe position $Y$ is denoted by $\nu_Y$.\\

To understand the motion of the probe when coupled quasistatically to its internal degree of freedom, we look at the statistical force $f(Y)$ on the probe. We will revisit that concept in Section \ref{statf}.  It is the mean force  when the probe is held fixed at position $Y$, obtained from averaging the mechanical force $-U'(Y,x) = -\id U(Y,x)/\id Y $ over $x$,
\begin{eqnarray}\label{1f}
f(Y) &=& -U'(Y,1) \,\nu_Y(1) - U'(Y,0) \nu_Y(0) - U'(Y,2)\nu_Y(2) \nonumber \\
&=& - U'(Y,0) + u'(Y) \,\nu_Y(1)  + v'(Y)\,\nu_Y(2)
\end{eqnarray}
for $\nu_Y(x), x=0,1,2$, the stationary occupation of the driven three-level system \eqref{rl}, and where
\[
u(Y) = U(Y,0)-U(Y,1) ,\qquad v(Y) =U(Y,0)-U(Y,2)
\]
The idea is that the probe (say of mass $M$) will (in infinite time-separation from its fast internal degree of freedom) have a dynamics of the form
\[
M\ddot{Y}_t = f(Y_t) - \gamma\,M\,\dot{Y}_t + \sqrt{2M\gamma k_BT} \xi_t, \qquad Y_t \in S^1
\]
where the friction $\gamma$ and the white noise $\xi_t$ are due to the thermal bath.
Whether a probe-current will develop depends on the presence of a rotational part in the force $f(Y)$.\\ 
We know from the previous subsection that the occupation law can be modified drastically by the activity parameters when $\ve \neq 0$, and hence we expect here also that the statistical force $f$ in \eqref{1f} will change significantly depending on the choice of $a(Y),b(Y),c(Y)$.  To illustrate that let us simplify to the case where $u(Y)=v(Y)$, in which case the rotational part of the force \eqref{1f} is
\begin{equation}\label{emf}
\text{emf} := \oint \id Y\,f(Y) = - \oint \id Y\, u'(Y) \,\nu_Y(0)
\end{equation}
Obviously, when $\ve=0$ (detailed balance) there is no current, whatever the $a(Y),b(Y),c(Y)$, because then $f(Y) = k_BT\,(\id/\id Y) \log \sum_x e^{-\beta U(Y,x)}$.  Next suppose that the activity parameters are ``thermodynamic'' in the sense that $a(Y),b(Y),c(Y)$ (as functions of $Y$) are (just) functions of $u(Y)$.  Then, clearly, also $\nu_Y(0) = G'(u(Y))$ for some function $G$, and emf $=0$ from \eqref{emf}.  It is therefore necessary that the activity parameters are (truly) kinetic for there to be a probe current (no matter how large the driving $\ve$).\\ 
To make everything more explicit we may compute the stationary distribution for $\ve \uparrow \infty$, and take $a(Y)=b(Y)=\delta(Y)\, c(Y)$ (keeping also $u(Y)=v(Y)$) for $\delta(Y) =O(\delta)$ of the order of some small $\delta$:  then surely,
\[
\nu_Y(0) = g_0(u(Y)) + \delta(Y) \, g_1(u(Y))  + O(\delta^2)
\]
and it will be the ``twist'' between the $Y-$dependence in $\delta(Y)$ {\it versus} in $u(Y)$, that will generate a probe-current via the emf \eqref{emf}; see more in \cite{nong}.
For a nonzero gradient force in \eqref{1f} we need not only that the (fast) three-level system is driven ($\ve\neq 0$) but also that there is some ``incommensurability'' between the activity parameters as function of the probe and the thermodynamic variables such as the energy and the entropy fluxes as function of the probe.  There are various ways to be mathematically precise about that incommensurability, but here we suffice by referring to \eqref{emf} to state that we need
$\oint \id Y\, u'(Y)\,g_1(u(Y)) \,\delta(Y) \neq 0$ to generate a probe current in order $\delta$.\\
\end{example}

\begin{example} [molecular motor]
Consider a dimer for a random walker, i.e., a walker occupying two neighboring sites $(x,x+1)$ on the integers, somewhat similar to a molecular motor like Myosin having two heads walking on an asymmetric substrate; see \cite{kel}.  There is also an upright position of the walker, where it stands up on one site.
There is no direct transition from one horizontal position $(x,x+1)$ to a next one, $(x+1,x+2)$ to the right or $(x-1,x)$ to the left. It must pass via an upright position at $x+1$, respectively at $x$ after which it falls flat again.  The transition rates are, for $x\in \mathbb{Z}$,
\begin{eqnarray*}
&&k((x,x+1) \rightarrow  x) =  a,\quad k((x,x+1) \rightarrow  x+1) =  b\\
 &&k(x \rightarrow  (x,x+1)) = k(x \rightarrow  (x-1,x)) = c
\end{eqnarray*}
The driving is more subtle here as it is the left-right asymmetry in the substrate $a\neq b$ that makes the reactivities for standing up at $x$ {\it versus} standing up at $x+1$ different when the dimer is flat on $(x,x+1)$. Clearly, the difference between $a$ and $b$ will make the dimer move.  If $a < b$, it will happen more often that the dimer tumbles to the right producing a current in that direction.\\
\end{example}

It is also essential to realize that the activity parameters may depend on external factors.   The driving amplitude could modify the reactivities. One reason is that the escape rates or the trapping behavior may very well change with external driving.  It implies that the currents may be steered to certain regimes.  We give the simplest example and how it relates to negative differential conductivities.\\

\begin{example}\label{rw}[inverse effect]
	There have by now been various studies concerning the phenomenon of ``getting more from pushing less'' \cite{zia}.  We come back to negative differential conductivities in Section \ref{smalln}, but here we give the simplest case.\\
	As every Markov jump process is essentially a random walker (possibly on a very complicated graph), we can as well start there and with the simplest architecture.  Let  $x\in \bbZ$ be the position of a walker jumping to nearest neighbor sites with rates $k(x,x\pm 1)=a(\beta,\epsilon)\,\exp[\pm \beta \epsilon/2]$.  The $\beta$ refers to an inverse temperature and $\epsilon$ is the external field.  For $\epsilon>0$ there is a bias to move to the right. The log-ratio $\log k(x,x + 1)/k(x+1,x)= \beta\epsilon$ is the entropy flux per $k_B$.  That is the Joule heat over temperature in the environment corresponding to the external work done on the walker to make it move one step to the right. Together with the activity parameter $a(\beta,\ep)$, the driving modifies the escape rate $\xi(x) = k(x,x+ 1)+ k(x,x- 1)= 2a(\beta,\epsilon) \,\cosh (\beta\epsilon/2)$. Similarly, the current (to the right) is influenced as
\[
J = 2a(\beta,\epsilon)\,\sinh(\beta\epsilon/2) 
\]	
In the linear regime for small external field, we have $J = a(\beta,0)\,\beta\,\epsilon$ with $\beta\,\epsilon \ll 1$.  That means that $a(\beta,0)$ relates to the diffusion constant in equilibrium; see more in Section \ref{mer}.  On the other hand, for large enough $\beta \epsilon$, the behavior of $J$ gets significantly influenced by $a(\beta,\epsilon)$.  That amplitude may for example start to decrease when $a(\beta,\epsilon)$ is sufficiently decreasing in $\epsilon$, which would mean that the current $J$ decreases under increased external field $\epsilon$ for large enough $\epsilon$.
In richer architectures, one can also observe current reversals; see \cite{springer}.
\end{example}

\subsection{Definition of frenesy}\label{defgre}
We next lift the idea of time-symmetric activity to path-space, just like \eqref{enp} is doing for the (time-antisymmetric) dissipation.  We still consider Markov jump processes as in \eqref{gfo}.  Trajectories, here piecewise constant, consist of either ``waiting'' or ``jumping'' events.  Between the jump times are intervals of waiting, the length of which is exponentially distributed with parameter given by the escape rate
\[
\xi(x) := \sum_y k(x,y)
\]
when in state $x$. If there is a density $\rho(x)$ over the states, then the expected escape rate is $ \frac 1{2}\sum_{x,y} \tau_\rho(x,y)$ in terms of the traffic
\begin{equation}\label{traff}
\tau_\rho(x,y) := \rho(x)k(x,y) + \rho(y)k(y,x)
\end{equation}  
That expected traffic over the bond $(x,y)$ is, in contrast with $j_\rho(x,y)$ in \eqref{dbc}, the symmetric non-directional current, a measure of dynamical activity as it gives the expected number of transitions/jumps (symmetrically) between $x$ and $y$ per unit time given the density $\rho$.  In fact, the traffic is a potential for the current in the sense that with \eqref{gfo},
\[
\frac{\partial \tau_\rho}{\partial s(x,y)} =  \frac 1{2} \,j_\rho(x,y)
\]
 Those escape rates obviously attend to hidden degrees of freedom and result from coarse graining.
If we have a trajectory $\omega = (x_s, 0\leq s\leq t)$ over the time $[0,t]$, then the time-integrated escape rate (path-wise escape) is
\begin{equation}\label{esc}
\text{Esc}(\omega) := \int_0^t\id s \,\xi(x_s)
\end{equation}
depending on the path $\omega$.  It is obviously symmetric under time-reversal: $\text{Esc}(\omega) = \text{Esc}(\theta \omega)$ for $(\theta\omega)_s = x_{t-s}, s\in [0,t]$.  In contrast, the path-wise entropy flux per $k_B$ of \eqref{enp}, $ S(\omega) = -S(\theta\omega)$, is antisymmetric under time-reversal.  Esc$(\omega)$ is nonzero even when there are no transitions at all in the path $\omega$, and {\it all is asleep}, \cite{pri}.\\
There is however also a time-symmetric component in the jumping or motion itself.  By being symmetric, the activity parameters \eqref{apa} contribute in the same way to the jump rates for the original as for the time-reversed trajectory.  For each path $\omega$ we therefore consider the activated traffic
\begin{equation}\label{dd}
\text{Act}(\omega) := \sum_s \log \frac{a(x_{s^-},x_s)}{a_0}
\end{equation}
where the sum is over all jump times in $\omega$ and $a_0$ is a (small but constant) reference frequency for the process.  When the $a(x,y)$ are constant, then $\text{Act} \sim {\cal T}$ where ${\cal T}(\omega)$ is the traffic, the total number of (unoriented) jumps in $\omega$.
The stationary expectation of \eqref{dd} per unit time is act$(\nu) = \frac 1{2}\sum_{x,y} \tau_\nu(x,y)\,\log a(x,y)/a_0$ in terms of the traffic \eqref{traff} again.
The activated traffic Act$(\omega)$ and the path-wise escape Esc$(\omega)$ are both time-symmetric but the first is related to jumps (hence, leaving the state) and the second is related to waiting (hence, being in the state).  We will see later in Section \ref{mdf} in what sense they can be considered a Legendre pair.\\

We call frenesy  associated to the path $\omega$, the quantity
\begin{eqnarray}
\label{fre}
D(\omega) &:=& \text{Esc}(\omega) - \text{Act}(\omega) \nonumber\\
&=& \int_0^t\id s\,\sum_y k(x_s,y)    - \sum_s \log a(x_{s^-},x_s)/a_0
\end{eqnarray}
$D$ is obviously time-symmetric, $D\theta=D$.\\
The expectation of the frenesy is a weighted sum over the local bond-traffic, 
\[
\langle D \rangle_\mu = \int_0^t \id s\,\sum_{x,y}\,\mu_s(x)
\,k(x,y)\,[1- \log a(x,y)/a_0]
\]
where $\mu_s$ solves the Master equation.   Its stationary value per unit time is
\[
\frac 1{t} \langle D \rangle = \frac 1{2}\sum_{x,y}\,\tau_\nu(x,y)\,[1- \log a(x,y)/a_0]
\]
in terms of the expected stationary traffic $\tau_\nu(x,y)$ between states $x,y$.\\
We are most often interested in changes in frenesy (more than in absolute values).  What matters are excesses or differences in frenesy, as we are used to also in the case of thermodynamic potentials.  We write the differences as $\Delta D = D - D_\text{ref}$ and $\Delta S = S - S_\text{ref}$ when we compare the frenesies and the entropy fluxes with respect to some (for the moment arbitrary) reference process.  For the entropy we refer to \eqref{enp}.  It is then easy to see that
  \begin{equation}\label{ahac}
  \frac{\text{Prob}[\omega]}{\text{Prob}_\text{ref}[\omega]} = e^{-{\cal A}(\omega)}, \quad {\cal A} = \Delta D - \frac 1{2} \Delta S
  \end{equation}
in which we introduced the action ${\cal A}$ that allows to move between dynamical ensembles.  If no confusion exists concerning the reference ensemble we abbreviate the trajectory weight simply as
  \begin{equation}\label{dyen}
  \text{Prob}[\omega] \sim e^{- D(\omega) + \frac 1{2}  S(\omega)}
  \end{equation}
  The above finally inspires a more general definition of frenesy.  {\bf We call excess in frenesy the time-symmetric part in the action for the relative weight of trajectories in the new process with respect to an original (reference) process.}\\

Below we give some other realizations of \eqref{ahac}--\eqref{dyen}.  Various techniques of deriving the action $\cal A$ and its decomposition in entropic and frenetic parts are contained in \cite{jmp2000} and belong in the general theory of stochastic calculus under the name of Cameron-Martin and Girsanov theorems.  The main point is to understand in explicit terms the change of measure, \cite{girs}.\\ 
Evidently, formula \eqref{dyen} has a meaning that goes far beyond its mathematical derivation (which we skip).  First, it wants to specify weights on spacetime trajectories, which is the ambition of a Lagrangian statistical mechanics.  Instead of concentrating on fixed-time distributions, we speak about dynamical ensembles. The formul{\ae} \eqref{ahac}--\eqref{dyen} prescribe how to make them.  Instead of specifying the nonequilibrium dynamics with Master equations or via Fokker-Planck equations, we collect the necessary information in two essential quantities, $S$ from (stochastic) thermodynamics (entropy fluxes under a condition of local detailed balance), and $D$ from kinetics (escape rates and traffic).  Taking the mechanical analogy, we may think of $S$ as the ``potential'' part (like the potential energy in a Lagrangian) and $D$ would then take the role of kinetic energy.\\
Secondly, dynamical ensembles with such weights are relevant to see the dynamical behavior and fluctuations in particular: the system takes such paths  as to maximize the entropy fluxes while minimizing the frenesy.  Dominant pathways may change from one reference case to another according to the dynamical principle summarized in \eqref{dyen}.  That can in fact already be applied usefully under the condition of
 detailed balance; see e.g. \cite{fac}.  In molecular dynamics simulation of glass formation, a similar frenetic order parameter is investigated in \cite{cha}.  In the following sections we will see more use of that (also) for (stationary) nonequilibria.
 In the event that the changed process does not modify the activity parameters $a(x,y)$, we only need to care about the change in the escape rates.  Then $\Delta D = \Delta$Esc, obtained from \eqref{fre}.  That corresponds also to the diffusion cases considered below.

 \subsubsection{Overdamped diffusion}
  A first case are overdamped diffusion processes, say in the case of a Brownian particle with position $x_t\in {\mathbb R}^n$,
 \begin{equation}
 \dot{x}_t = \chi \cdot F(x_t) + \sqrt{2k_BT\,\chi}\,\xi_t,\qquad \xi_t=\mbox{ standard white noise}\label{overd}
 \end{equation}
The mobility $\chi$ is a positive definite $n\times n$-matrix not depending on $x$ (for simplicity only) in front of the total force given by
\[
F(x) = h\,f(x)  + g(x)
\]
We want the excess frenesy and entropy flux per $k_B$, for $h\neq 0$ with respect to the (reference) dynamics with $h=0$.  The result is
\begin{eqnarray}\label{frda}
  D(\omega) &=& \frac{h^2\beta}{2}\int_0^t \id s\, f\cdot \chi f + h\beta\int_0^t \id s \,f\cdot \chi\,g + h \int_0^t \id s\, \chi\nabla\cdot f\\
S(\omega) &=& h\,\beta\int_0^t \id x_s \circ f(x_s)\nonumber
\end{eqnarray}
The stochastic integral with the $\circ$ is in the sense of Stratonovich.  We see that here the highest order is quadratic in the excess parameter $h$.  We can also specify to the case where instead, in \eqref{overd},
 \[
 F = \epsilon f - \nabla U \label{force}
 \]
 for a potential $U$. Then, $f$ stands for the nonconservative (or rotational) part of the force $F$ with strength $\epsilon$.
The reference dynamics satisfies the condition of detailed balance (time-reversibility) when $\epsilon = 0$. 
The excess frenesy in taking $0\rightarrow \epsilon$ equals
\begin{equation}\label{fred}
  D(\omega) = \frac{\epsilon^2\beta}{2}\int_0^t \id s\, f\cdot \chi f - \epsilon\beta\int_0^t \id s \,f\cdot \chi\nabla U + \epsilon \int_0^t \id s\, \chi\nabla\cdot f
\end{equation}
The entropy flux per $k_B$ remains $\beta$ times the work done by the nonconservative
 force.\\

Perhaps the expression \eqref{fred} does not speak totally by itself.  Then it is good to see how it is obtained by a limiting procedure starting from a jump process.\\
Consider a lattice mesh $\delta>0$ for $x\in \delta \mathbb{Z}$, and a walker with detailed balance transition rates
 \[
 k_o(x,x\pm\delta) = {\cal D}\,\exp-\frac{\beta}{2}[U(x\pm \delta) - U(x)],\;\;\quad {\cal D}>0
 \]
We introduce driving in new rates given by
 \[
 k(x,x\pm\delta) = k_o(x,x\pm\delta)\,\exp \big[ \frac{\beta \epsilon \delta}{2}f(x,x\pm\delta) \big]
 \]
 Since the activity parameters do not change with $\epsilon$, we only have to see here for the change in escape rates and we must calculate the excesses
 \[
 k(x,x+\delta)-k_o(x,x+\delta) + k(x,x-\delta) - k_o(x,x-\delta)
 \]
 which to order $\delta^2$ equal
 \begin{eqnarray}
 && {\cal D}\,(1-\frac{\beta \delta}{2} U'(x))\,\big[\frac{\beta \epsilon \delta}{2}\,f(x,x+\delta) + \frac{\beta^2 \epsilon^2 \delta^2}{8} f^2(x,x+\delta)\big] \nonumber\\
 +&& {\cal D}\,(1+\frac{\beta \delta}{2} U'(x))\,\big[\frac{\beta \epsilon \delta}{2}\,f(x,x-\delta) + \frac{\beta^2 \epsilon^2 \delta^2}{8} f^2(x,x-\delta)\big] \nonumber\\
 =&& \frac{{\cal D}\beta^2\epsilon^2 \delta^2}{8}\,[f^2(x,x+\delta) + f^2(x,x-\delta)] + \frac{{\cal D}\beta \epsilon \delta}{2}\,[f(x,x+\delta) - f(x-\delta,x)] \nonumber\\
 &-& \frac{{\cal D}\beta^2 \delta^2\epsilon}{4}U'(x)\,[f(x,x+\delta) + f(x-\delta,x)]\nonumber
 \end{eqnarray}
Finally, setting ${\cal D}\beta=\chi$ and $f(x,x)=f(x)$,
 \begin{eqnarray}
 &&\lim_{\delta\downarrow 0} \frac{2}{\delta^{2}}[k(x,x+\delta)-k_o(x,x+\delta) + k(x,x-\delta) - k_o(x,x-\delta)]=\nonumber \\
 =&&\frac{\chi \beta}{2} \epsilon^2 f^2(x) + \chi\epsilon f'(x) -\chi\beta\epsilon f(x) U'(x)
 \end{eqnarray}
which we now recognize in \eqref{fred} (in one dimension).

 \subsubsection{Underdamped diffusion}
 Next we take the example of a Langevin dynamics for a particle with mass $m$, position $q_t$ and velocity $v_t$, in one-dimensional notation,
 \begin{eqnarray}
  \dot{q}_t &=& v_t  \nonumber\\
  m\dot{v}_t &=& [F(q_t)- m\gamma v_t] + \sqrt{2{\cal D}}\,\xi_t\label{underd}
 \end{eqnarray}
where $\gamma$ is the constant friction and $\xi_t$ is always standard white process.  The (symmetric matrix, in general) ${\cal D}$ governs the variance of that noise.
 The Einstein relation between $\gamma$ and ${\cal D}$ introduces the inverse temperature $\beta$ of the environment: $\gamma = \beta {\cal D}$.  The nonequilibrium resides possibly in the 
  force $F = \epsilon f - \nabla_q U$ where $f$ may correspond to a bulk driving.  The (generalized) detailed balance case has
  $\epsilon = 0$ but the time-reversal transformation must include the flipping of the velocities along the trajectory.\\
  We compute the action for the change of process when $0\rightarrow \epsilon$.  We have
(again without putting $\Delta D$ and $\Delta S$ for the differences)
 \[
 \frac{\text{Prob}_{\epsilon}(\omega)}{\text{Prob}_0(\omega)} = \exp \left[ -D(\omega) + \frac{S(\omega)}{2}\right]
\]
 with
 \begin{eqnarray}\label{smo}
D(\omega) &=& \frac{\epsilon^2}{2}\int_0^t \id s f\cdot {\cal D}^{-1} f
                          - \epsilon\int_0^t \id s \,f\cdot {\cal D}^{-1}\nabla U - m\,\epsilon \int_0^t dv_s \circ {\cal D}^{-1} f\\
  S(\omega) &=& \epsilon\beta\int_0^t \id s \,v \cdot f
 \end{eqnarray}
 $S$ equals the work done by the nonconservative force $f$, times $\beta$.  The frenesy $D$ consists of several terms related to the kinetics, as explained for \eqref{fred}.\\

 For a case of boundary driven underdamped dynamics we can check e.g. the heat conduction networks in \cite{heatcond}.
  We consider the process
 $P_{\beta}^{\kappa}$ corresponding to the underdamped dynamics,
 \begin{eqnarray}\label{eqns1}
 \dot q_{i}&=&p_{i}, \quad i\in V\nonumber\\
 \dot p_{i}&=&-\frac{\partial U}{\partial q_{i}}(q), \quad i \in V\setminus \partial V\\
 \dot p_{i}&=&-\frac{\partial U}{\partial q_{i}}(q)-\gamma\ka_{i}
 p_{i} +\sqrt{\frac{2\gamma}{\be_i}}\,\xi_i(t), \quad i\in \partial
 V\nonumber
 \end{eqnarray}
 Here $q_i,p_i$ are respectively the position and the momentum at site $i$, both scalars.  The ``volume'' $V$ consists of the vertices of a graph, thought to be a piece of crystal lattice.  That volume has a boundary $\partial V$ (some selection of sites) to which the various equilibrium heat baths are coupled.  There, at those boundary sites, Langevin forces are added with friction $\gamma \kappa_i$ at $i\in \partial V$ and noise amplitude $\gamma/\beta_i$; the rest of the dynamics is Hamiltonian with potential $U$ being e.g. a sum over interaction potentials defined over the bonds of the graph. We see there is generalized detailed balance when  the $\ka_{i}\be_i=\beta$, $\forall i\in \partial V$, corresponding to a unique environment temperature $\beta^{-1}$.  Then indeed, 
 the process (\ref{eqns1}) is generalized reversible with respect to the Maxwell--Boltzmann distribution $\nu_\text{eq}(q,p) \propto \exp -\beta H$, for $H= \sum_{i \in V} p_i^2/2 + U(q)$ where $U$ may consist of a self-energy and some local interaction: when the $\ka_{i}\be_i=\beta$ for all boundary sites, then
 $P_{\beta}^{\kappa}=P_{\beta}^{\kappa} \Theta$
 where $(\Theta \omega)_s = (-p_{t-s},q_{t-s})$ is the time-reversal on phase-space trajectories $\omega =
 ((p_s,q_s), s\in [0,t])$.\\

The evolution \eqref{eqns1} becomes a nonequilibrium dynamics for $\kappa_i\equiv 1$.  Then, we really have different temperatures $\sim \beta_i^{-1}$ at the boundary sites.  There is a corresponding (excess) frenesy (in going from $\kappa_i = \beta/\beta_i \rightarrow \kappa_i = 1)$ which is equal to
 \begin{equation}\label{aaa}
 D(\omega) = -\sum_{i\in\partial V}\left(\frac{1}{2}
 (\beta-\be_i)[p_i(t)-p_i(0)]
+ \gamma\,\frac{\beta^2 -\beta_i^2}{\beta_i} \,\int^t_0 p^2_i(t)\,\id t\right)
  \end{equation}
Apparently, its time-extensive part is directly related to a weighted time-integrated kinetic energy at the boundary, as measure of dynamical activity.  We can see indeed how the activated traffic is related to kinetic energy; see also Section \ref{kinu}.\\

\subsubsection{Time-dependent diffusion}
Let us also add a time-dependent example. Consider the following dynamics (in one-dimensional notation for simplicity) for position $x_t$ and velocity $v_t$ of a probe in a thermal medium where the external force changes with time:
\[
\dot x_t = v_t
\]
\[\dot v_t = -\gamma\,v_t + F(x_t,\lambda_t) + \sqrt{2\gamma T} \,\xi_t
\]
where we put the mass equal to one, $\gamma>0$ is a damping and $T$ is the temperature of the environment.  The time-dependence of the force $F$ is governed by a protocol for the parameter $\lambda_t$.\\
As a reference process we take the homogeneous Langevin dynamics
\[
\dot x_t = v_t
\]
\[\dot v_t = -\gamma\,v_t  + \sqrt{2\gamma T} \,\xi_t
\]
without external force.\\ 
The time-dependent process $P$ can be expressed with respect to the reference process
\[
P(\omega) = e^{-{\cal A}(\omega,\lambda)}\,P_\text{ref}(\omega)
\]
for trajectories $\omega = ((x_s,v_s), 0\leq s\leq t)$, in terms of an action ${\cal A}$ with
\[
4\gamma T\,{\cal A}(\omega,\lambda) = \int_0^t \id s\left[ (\dot{v}_s + \gamma v_s - F(x_s,\lambda_s))^2 -(\dot{v}_s + \gamma v_s)^2 \right]
\]
\begin{equation}\label{ac}
	= \int_0^t \id s\left[ -2\dot{v}_sF(x_s,\lambda_s) - 2\gamma v_sF(x_s,\lambda_s) + F(x_s,\lambda_s)^2 \right]
\end{equation}

In the above generality there is only one immediate symmetry transformation, which is time-reversal $\theta$ under which $\theta(x_s,v_s)= (x_{t-s},-v_{t-s})$. For the protocol,  we reverse it via $(\theta\lambda)_{s} =  \lambda_{t-s}$.\\
Then, we have from \eqref{ac},
\[
S(\omega) = {\cal A}(\theta\omega,\theta \lambda)  - {\cal A}(\omega,\lambda) =
\frac 1{T} \int_0^t \id s\, v_s\,F(x_s,\lambda_s)
\]
which we recognize as the time-integrated power divided by temperature, which is instantly dissipated as Joule heat in the environment. That gives the entropy change $S$ in the environment per $k_B$, as function of the particle trajectory $\omega$ and for given protocol $\lambda_s, 0\leq s\leq t$.\\

For the frenesy $D = ({\cal A}\theta + {\cal A})/2$ we find
\[
D(\omega) =  \frac 1{4\gamma T}\,\int_0^t \id s\left[ F(x_s,\lambda_s)^2  -2\,\dot{v}_sF(x_s,\lambda_s)\right]
\]
where the first term refers to an escape rate and the second term (with Stratonovich integral) to the traffic.   Note that it depends also on the acceleration $\dot{v}_s$.  For small $\gamma$ and $ T$, $\gamma T\ll vF$, we see however that $D(\omega) \simeq -\frac 1{4\gamma T}\,\int_0^t \id s\, F(x_s,\lambda_s)^2$ entirely in terms of the force.\\
For the harmonic oscillator 
\[
F(x_t,\lambda_t) = -\lambda_t x_t
\]
with changing spring constant $\lambda_t$ we get the frenesy in the form
\[
D_\text{ho}(\omega) =  \frac 1{4\gamma T}\,\int_0^t \id s\left[\lambda_s^2\,x_s^2  +2\,\lambda_s\,\dot{v}_s\, x_s\right]
\]
which is relevant for an underdamped particle in a time-dependent harmonic trap.  Similar expressions are easily derived in the overdamped case as well.\\

\subsubsection{Boundary driven Kawasaki process}
To end we go back to Markov jump processes and we give the frenesy for an interacting boundary driven Kawasaki dynamics.\\
We consider the interval $\{1,2,\ldots,N\}$ of sites $i$ that carry at most one particle, $x(i)=0, 1$  (vacant or occupied).  The bulk dynamics is governed by a Hamiltonian
\[
H(x) = \kappa\sum_{i=1}^{N-1} x(i)x(i+1) - \mu\sum_{i=1}^N x(i) 
\]
giving here the energy of a system of indistinguishable particles making configuration $x= (x(i),i=1,\ldots,N)\in \{0,1\}^N$.
We work with bulk entropy fluxes $s(x,y) = (H(x)-H(y))/{k_BT}$ to an equilibrium environment at temperature $T$. The bulk transitions $x\rightarrow y$ are particle hopping to a nearest neighbor site if that one is empty.   At the two edges (left and right) particles can enter or leave the system.  That is summarized in the usual transition rates
\[
k(x,y) = a(x,y)\, e^{[H(x)-H(y)]/(2k_BT)}  
\]
except for the left boundary where, when $x=y$ except for $y(1) = 1 - x(1)$, we make
\begin{equation}\label{mul}
k(x,y) = \left(\,e^{-\delta\beta/2} x(1) + e^{\delta\beta/2} (1-x(1))\,\right)\,a(x,y) \,e^{H(x)-H(y)/(2k_BT)}
 \end{equation}
It is easy to verify that whenever $\delta=0$ in \eqref{mul}, then the dynamics satisfies the condition of detailed balance with respect to the grand canonical Gibbs distribution (at inverse temperature $\beta$ and with chemical potential $\mu$), in the sense that for all particle configurations $x,y$,
\[
k(x,y)\, \exp[-\beta H(x)] = k(y,x)\, \exp[-\beta H(y)]
\]
For $\delta\neq 0$ detailed balance is broken and the left edge gets connected with a chemical reservoir with chemical potential $\mu+\delta$. In that case the stationary distribution is very complicated and in general unknown, but we know there will then appear a stationary particle current from left to right when $\delta>0$. Since we let $a(x,y)$ not to depend on $\delta$, the excess frenesy is
only due to the change in escape rates,
\begin{equation}\label{del}
D(\omega) =  \int_0^t\id s\left( x_s(1) [e^{-\delta\beta/2}-1] +  (1-x_s(1)) [e^{\delta\beta/2}-1] \right) 
\end{equation}
and, in $\omega$, depends only on the total occupation time at the left edge.  Observe that there is no trace of the particle interaction in that time-symmetric path-variable.

\begin{remark}
	The emphasis on dynamical ensemble and the approach of presenting the action does not need to start from Markov dynamics (as was the case above). Lagrangian statistical mechanics and model building can also depart directly from giving the action in terms of an entropic and frenetic part even without explicitly referring to a dynamics. That is not incompatible with approaches that go under the name of {\it maximum entropy} or {\it maximum caliber} principle, but only when frenesy is taken into account might there be the hope to obtain an ensemble relevant for nonequilibrium physics also away from close-to-equilibrium regimes.
\end{remark}
  
\begin{remark}
	That the time-symmetric and kinetic features of path-space distributions matter is not surprising of course, and also not original.  In the past people have been observing that non-thermodynamic aspects truly matter for nonequilibrium purposes, \cite{land1975}, and books have been written about e.g. the essence of quiescence in biological systems, \cite{hade}.  Obviously, dynamical activity has been discussed at length in recent years in contexts of trapping models, or dynamical phase transitions, or when there are kinetic constraints.  In particular it has given rise to a new paradigm for glassy dynamics.  The glass transition becomes an order--disorder phenomenon in space–time, with dynamical activity as order parameter \cite{mer}.  More on large deviations of spacetime trajectories and on dynamical phase transitions was then treated in \cite{viv,gar2,gar3}.\\   
	The more general and earliest push, at least for the author, to be systematic in nonequilibrium steady states about that time-symmetric part has come from response theory.  As we discussed in \cite{maarten} the fluctuation symmetry for the entropy production, while non-perturbative and giving many identities, as such fails to give information about the time-symmetric sector. From then on, a mixed terminology was used with words like ``activity'' and ``traffic'' dominating the discussions.  The word ``frenetic'' first appeared then in \cite{fdr}. Still later we settled for the word ``frenesy'' as it sounds similar to ``entropy'' or ``energy'', but also connects with its meaning as ``frequency.''  Moreover, it is directly related to the words ``frenzy'' and ``frantic,'' and originates from the Greek $\phi \rho \acute{\eta} \nu$ (phren), which probably comes closest to its precise meaning.  
\end{remark}  
  
\section{Dynamical fluctuations}\label{dynf}

Frenesy speaks about the time-symmetric fluctuation sector.  Obviously, that mixes with the antisymmetric sector in various ways but the frenesy adds very specific aspects.

\subsection{Symmetries}
The way the action is decomposed in \eqref{dyen}, in terms of a time-symmetric and a time-antisymmetric part, is obviously the most immediate relevant aspect for understanding trajectories in nonequilibrium physics.  Yet, depending on the situation, more and other symmetries will add to decompose the various parts even further or to identify sources of other symmetry-breaking.  We refer for the following to \cite{maarten,wojciech} for more examples and corresponding additional fluctuation symmetries.  Large deviations of time-symmetric activities have for example been considered in \cite{viv,rold,carlo}.\\

Denoting by $P(\omega)$ the probability of a path $\omega$ in the original process, we write $P_m(\omega)$ for  the probability of the same path in the modified process.  As always it is assumed that the modification is compatible with the original process in the sense that the same paths are either allowed or not allowed: $P_m(\omega)=0$ if and only if $P(\omega)=0$.  Then, when starting from the same initial state, we follow \eqref{dyen} to write
\begin{equation}\label{act}
P_m(\omega) = P(\omega)\, \exp[- D(\omega) + \frac 1{2} S(\omega)]
\end{equation}
 Writing $\theta$ for the time-reversal transformation on paths, there is always 
\[
\frac{P_m(\omega)}{P_m(\theta \omega)} = \frac{P(\omega)}{P(\theta \omega)}\, \exp[ S(\omega)]
\]
which gives rise to amply discussed fluctuation symmetries for $S$ and related quantities, \cite{poincare}.  On the other hand, if there is also (another) involution $\cal I$ on path-space for which $S= S\cal I$ is symmetric, then
\begin{equation}\label{df}
\frac{P_m(\omega)}{P_m(\cal I \omega)} = \frac{P(\omega)}{P(\cal I \omega)}\, \exp[ D(\cal I\omega) - D(\omega)]
\end{equation}
and the part $D\cal I - D$ of the (excess) frenesy which is antisymmetric under $\cal I$ will satisfy similar fluctuation symmetries as the variable entropy flux $S$ is famous for.  For example, from \eqref{df}, when the original process is $\cal I-$symmetric, $P(\omega)= P(\cal I \omega)$, we get\\
\begin{equation}\label{dfa}
1 = \int \id P_m(\omega)\, \exp[ D(\omega) - D(\cal I\omega)] = \left< \exp(\, D - D\,\cal I\,) \right>_m
\end{equation}
which is formally similar to the Jarzynski identity; see \cite{jar} for a review of the entropic versions.  Other symmetries are easily derived including the asymptotic ones in time but it is of course depending on the specific physics what kinetic variables are picked up by the difference $D-D\cal I$.  An interesting choice is to take $\cal I=$ spatial reflection, charge conjugation or mirror symmetry composed with time-reversal.  The usual arguments lead then for example to the fact that there is more traffic near the high temperature or high concentration boundaries. The argument starts from \eqref{dfa} via Jensen inequality to give
\begin{equation}\label{5l}
\langle D\cal I -D\rangle_m \geq 0
\end{equation}
To be more specific, let us revisit \eqref{del} and take $\cal I$ as the involution that flips the parameter $\delta$ (which can be considered a dynamical variable even though it is constant) composed with time-reversal.  Then, $D\cal I -D = 2\sinh (\delta\beta/2)\int_0^t\id s\,(2x_s(1)-1)$, and we conclude from \eqref{5l} that when $\delta>0$, then the time that the left edge was occupied must be larger than $t/2$. In the stationary distribution of the modified process, the density near the reservoir with the higher chemical potential must be larger.  Such a conclusion is not surprising but observe that the statement is very general and nonperturbative; no other derivations with such power appear to be around.  Similar general conclusions can be drawn for other models; e.g., from \eqref{aaa} and for two temperatures (at the boundary sites of a lattice interval) we see that the time averaged kinetic energy at the higher-temperature boundary must be bigger than at the lower-temperature boundary etc.\\

There is another important lesson that can be drawn from \eqref{df}.  Suppose indeed there was {\it no} frenesy; put $D=0$ in the action of dynamical ensembles.  Then, whenever the reference shows $\cal I$-invariance, $P(\omega) = P(\cal I \omega)$, \eqref{df} implies
\begin{equation}\label{dfah}
P_m(\omega) = P_m(\cal I \omega)
\end{equation}
and the $\cal I-$symmetry can never be broken.  The transformation $\cal I$ could for example be a composition of time-reversal and spatial reflection, for which \eqref{dfah} would predict that time-reversal on trajectories is the same as spatially reflecting them.  That that is not true when far-from-equilibrium is especially clear for turbulent flow around obstacles, or from asymmetric dune formation in aeolian sand transport, \cite{maarten,kroy}: reflecting the thermodynamic force at the same time as the arrow of time may leave invariant the entropy flux $S=S\cal I$, but not the frenesy $D\neq D\cal I$, and that shows itself far enough from equilibrium by deviations from laminar flow or from the creation of spatial asymmetries.\\

There is also a fluctuation relation jointly with the variable entropy production.  Upon integrating \eqref{act} over all paths $\omega,$ we get the normalization of the left-hand side,
  \begin{equation}\label{norma}
 \left< e^{S/2-D} |x_0=x\right>_o = 1
\end{equation}
with expectation $\langle \cdot \,|\,x_0=x\,\rangle_o$ under the reference dynamics $P$ and started from (an arbitrary state) $x$. One choice is to take $P$ to be an equilibrium reference process. That identity \eqref{norma} can then itself be expanded in order of nonequilibrium driving as present in $S$ and $D$.  For example, one will see that in linear order the expected entropy flux and the excess dynamical activity are equal; it thus appears that close-to-equilibrium entropy flux and frenesy merge in expectation.  New nonequilibrium effects due to a nontrivial frenetic contribution may only start at second order around equilibrium.  We refer to the recent \cite{rold} for more frenetic symmetries.


\subsection{Macroscopic dynamical fluctuations}\label{mdf}
In a macroscopic dynamical fluctuation theory, \cite{com,revjona,jona} we examine the probability of empirically possible trajectories.  Pioneered in \cite{O,OM} one tries to understand the structure of fluctuations as inherited from microscopic constituents and laws, and as relevant for typical macroscopic behavior, \cite{spohn,gen1}.  Mathematical references include \cite{DV,fen}, spanning almost half a century.\\
  Trajectories are specified at times $s\in [0,t]$ by giving the density $\rho_s$ and the current $j_s$.   For the density $\rho_s$ at time $s$ we may keep in mind the mass or energy density fields, or $\rho_s$ can also be an empirical average of particle properties at time $s$. The density changes in time and the empirical rate of change is governed by the physical current $j_s$.  There is always a constraint of the form 
\begin{equation}\label{conj}
\dot{\rho_s} + {\cal \nabla} j_{\rho_s} = 0,\qquad s\geq 0
\end{equation}
 where $\cal \nabla$ is a divergence (like in the continuity equation) or some other operator when the $\rho_s$ does not correspond to a locally conserved quantity.  Depending on the microscopic laws, constituents, level of coarse graining and initial conditions we want to know the probability of a \emph{possible} density-current trajectory $(\rho_s,j_s), s\in [0,t]$.    One must keep in mind a macroscopic limit in terms of a scale parameter $N \uparrow \infty$, like the number of independent copies of a finite Markov process, or the number of particles, the volume, {\it etc.} 
When the density field is well-chosen, so as to give rise to an autonomous evolution (first order in time) in the macroscopic limit, the asymptotic exponential probability of such a trajectory is given as
\begin{equation}\label{prob}
\text{Prob}[(\rho_s,j_s), 0\leq s\leq t] \simeq e^{-N {\cal F}(\rho_0)}\,e^{-N\int_0^t\id s\, {\cal L}(\rho_s,j_s)}
\end{equation}
Initially, some density $\rho_0$ appears with some probability.
When that probability refers to thermodynamic equilibrium at temperature $T$, then the functional ${\cal F}$ in \eqref{prob} is a thermodynamic potential (typically in units of $k_BT$) giving the statistics of the density at time zero.  The Lagrangian ${\cal L}$ in \eqref{prob} governs the dynamical fluctuations in determining the plausibility of the various possible trajectories.  Obviously, it also depends on all present forces and possible driving; ${\cal L}(\rho,j) = {\cal L}(\rho,j;F)$.\\
We can assume that ${\cal L}(\rho,j)\geq 0$ and that it is convex in $j$ for all $\rho$. Observe that for large $N$ the weights in \eqref{prob} are exponentially small except for the zero--cost flow $j_s^* = j^*(\rho_s)$ which solves ${\cal L}(\rho_s,j_s^*)=0$ for all times $s$.  In that sense, the zero-cost flow is overwhelmingly more probable in the macroscopic limit, and coincides with the hydrodynamic limit.

Similar to the ideas around \eqref{dyen} we decompose the Lagrangian ${\cal L}$ in a time-symmetric and a time-antisymmetric part,
\begin{equation}\label{ep}
{\cal L}(\rho,j) = \frac 1{2}[{\cal L}(\rho,j) + {\cal L}(\rho,-j)] - \frac 1{2}\sigma(\rho,j)
\end{equation}
where $\sigma(\rho,j) = {\cal L}(\rho,-j) - {\cal L}(\rho,j)$ is antisymmetric under time-reversal.  As before, under the physical condition of local detailed balance, that $\sigma(\rho,j)$ is the entropy production rate per $k_B$ corresponding to the couple $(\rho,j)$. Working under that local detailed balance we put $\sigma(\rho,j) = F\cdot j$, linear in $j$ with thermodynamic force $F= F(\rho)$.\\  For the time-symmetric part of the Lagrangian, we follow \eqref{fre}:
\begin{equation}\label{tsc}
{\cal L}(\rho,j) + {\cal L}(\rho,-j) = 2\psi(\rho,j) + 2{\cal L}(\rho,0)
\end{equation}
The ${\cal L}(\rho,0)\geq 0$ is the escape rate from condition $\rho$,
\[ 
\text{Prob}[(\rho_s\equiv \rho,j_s\equiv 0), 0\leq s\leq t] \simeq e^{-N {\cal F}(\rho)}\,e^{-N\,t\,\, {\cal L}(\rho,0)}
\] 
while $\psi(\rho,j) = \psi(\rho,-j) $ in \eqref{tsc} gives the activated traffic corresponding to a current $j\neq 0$.
Combining \eqref{ep}--\eqref{tsc} with $\sigma(\rho,j) = F \cdot j$ yields
\begin{equation}\label{ps}
\psi(\rho,j) = {\cal L}(\rho,j) - {\cal L}(\rho,0) + \frac 1{2}F\cdot j
\end{equation}
  Then,  $\psi(\rho,j)$ is convex in $j$, $\psi(\rho,0)=0$ and $\psi(\rho,j)=\psi(\rho,-j)$, which implies that $\psi(\rho,j)\geq 0$. Furthermore, from \eqref{ps} we can check that $\psi(\rho,j)$ and ${\cal L}(\rho,0)$ are related by Legendre transform: 
\[
2\,{\cal L}(\rho,0) = \sup_j\,\{F\cdot j - 2\psi(\rho,j)\}
\]
Per consequence, $0\leq {\cal L}(\rho,0) = \widehat\psi(\rho,F/2)$ is the Legendre transform of $\psi(\rho,\cdot)$ at force $F/2$.  Therefore, the decomposition \eqref{tsc} is in a pair of convex duals, with current and force being conjugate.  The Lagrangian \eqref{ep} thus obtains the canonical structure
\begin{equation}\label{eps}
{\cal L}(\rho,j) = \psi(\rho,j) + \widehat\psi(\rho,F/2) - \frac 1{2} F\cdot j
\end{equation}
Such a structure of joint density-current fluctuations away from equilibrium was developed in \cite{lag1,lag2,epl,MPR13,pons} to which we refer for more discussion.  The macroscopic frenesy corresponds to the first two terms in \eqref{eps}, similar to \eqref{fre}.  Here however, the Lagrangian ${\cal L}(\rho,j) \geq 0$ always.  Note also that when the Lagrangian equals zero, the frenetic and the entropic part must coincide, \cite{fren,ncwh}:
\[
\psi(\rho,j^*) + \widehat\psi(\rho,F(\rho)/2) = \frac 1{2}\,F(\rho)\cdot j^*
\]
which determines the zero cost flow $j^* = j^*(\rho)$ at density $\rho$, and the hydrodynamic equation when inserted in \eqref{conj}.\\

The above can of course be made explicit for a number of processes; we need so called dynamical large deviations.  The simplest case to consider are many ($N$) independent copies of the same Markov process.  We refer to \cite{lag1}  for the case of $N$ independent jump processes, for which we here summarize the main resulting structure.\\
The computed Lagrangian is $\caL(\rho,j) = \bbL(\rho,k^j)$ with
\[
\bbL(\rho,\lambda) = \sum_{x, y \neq x} \rho(x) \Bigl[ \lambda(x,y) \log \frac{\lambda(x,y)}{k(x,y)} - \lambda(x,y) +
  k(x,y) \Bigr]
  \]
and
\begin{equation}\label{eq: j-explicit}
  k^j(x,y) = \frac{1}{2\rho(x)}\{j(x,y)
  + [j^2(x,y) + 4 \rho(x) \rho(y) k(x,y) k(y,x)]^{\frac{1}{2}}\}
\end{equation}
That is explicit in terms of the original rates $k(x,y)$.  We can also linearize that Lagrangian $\caL(\rho,j)$ around the steady condition, even when far-from-equilibrium.
To examine the structure of small deviations around that steady nonequilibrium condition we define the rescaled Lagrangian
\begin{equation}\label{eq: scaled}
  L(p,\frj;F) = \lim_{\ep\downarrow 0} \ep^{-2}
  \caL(\nu (1+ \ep\,p), \bar j + \ep \frj; F)
\end{equation}
Here, $\bar j(x,y) = \nu(x) k(x,y) - \nu(y) k(y,x)$.
We have indicated that the driving $F$ is kept fixed while both the density
and the current are expanded around the stationary values $\nu$,
respectively $\bar j$.  We will use the
notation
\begin{equation}\label{eq: traffic}
  \bar \tau(x,y) = \nu(x) k(x,y) + \nu(y) k(y,x)
\end{equation}
for the steady traffic; see Section \ref{defgre}.  The scaled Lagrangian~\eqref{eq: scaled} is computed from \eqref{eq: j-explicit} in \cite{lag1} to be
\begin{equation}\label{la}
  L(p,\frj;F) = \sum_{(xy)} \frac{1}{4\bar \tau(x,y)}\,\bigl\{2 \frj(x,y) - \bar \tau(x,y)\, [p(x) - p(y)]
  - \bar j(x,y)\, [p(x)+ p(y)] \bigr\}^2
\end{equation}
That is the Lagrangian describing normal fluctuations around the steady nonequilibrium condition.
The steady traffic $\bar \tau$ plays the role of a variance
in this fluctuation law.\\
A special and known case is the Onsager-Machlup situation \cite{OM}, where on looks at the small fluctuations around an equilibrium condition where $\nu=\nu_\text{eq}$ and $\bar j =0$.
If we would have started with a detailed balance dynamics with transition rates 
$k(x,y) = \nu_\text{eq}^{-1}(x)\,\ga(x,y)$, $\gamma(x,y)=\gamma(y,x)$, the scaled
Lagrangian~\eqref{eq: scaled} with $F=0, \nu=\nu_\text{eq}, \bar j =0$ is 
\begin{equation}\label{ba}
  L_\text{eq}(p,\frj;F=0) = \sum_{(xy)} \frac{1}{4\ga(x,y)}
  \bigl\{\frj(x,y) - \ga(x,y) [p(x) - p(y) ] \bigr\}^2
\end{equation}
From comparing \eqref{la} with \eqref{ba} we observe that the structure of small
fluctuations (albeit both Gaussian by construction) is remarkably different when moving away from equilibrium.  Observe indeed that in the stationary regime
$\sum_{y \neq x} \frj(x,y) = 0$ so that the empirical distributions
of occupations and of currents become uncorrelated  around equilibrium, $\sum_{(xy)}\frj(x,y)[p(x)-p(y)]=0$.   In nonequilibrium there emerges an occupation-current coupling
proportional to $\bar j/\bar{\tau}$.

\subsection{Frenetic variational principle}

One sometimes wonders whether there also exist variational principles in nonequilibrium statistical mechanics.  The answer is positive, but the relevant issue is the extent to which those principles are useful and operational.  In equilibrium the stationary (equilibrium) distributions are determined by an energy--entropy balance, e.g. thermal equilibrium is the condition that minimizes the (suitable) free energy, which is the celebrated and extremely powerful Gibbs variational principle.  Similarly but on spacetime we have \eqref{dyen} to determine the dynamical ensemble, and we might as well ask for a variational principle that starts from there.\\
Here is then the idea in the framework of Markov jump processes as from \eqref{gfo}, \eqref{enp}, \eqref{fre}.  We want to characterize {\it variationally} the stationary distribution $\nu$ for a jump process with transition rates $k(x,y)$.  Suppose we change those transition rates by adding a potential $V$ in the form
\begin{equation}\label{mad}
k_V(x,y) = k(x,y) e^{V(x)-V(y)}
\end{equation}
We compare the two dynamical ensembles, $P^V$ and $P$.
For the action of $P^V$ with respect to the original process $P$, we proceed as in \eqref{dyen}.  Their initial conditions are taken identical so that
\[
\log \frac{P^V}{P} = D-D^V + \frac{S^V-S}{2}
	\]
in terms of excesses in frenesy and in entropy flux. 
If we take the average over that expression with respect to the $P^V$-process we get the relative entropy (also called Kullback-Leibler divergence) of $P^V$ with respect to $P$.  Note however that $S^V-S$ is a time-difference $V(x_t)- V(x_0)$.  The change in entropy flux is therefore negligible as we add a total difference, which is not time-extensive.  The big change is in the frenesy, more specifically in the escape rates. We can mathematically estimate the relative entropy after the new escape channel was opened.
The new process $P^V$  over the time $[0,t]$  has distribution $\rho_s$ at time $s$, solving the Master equation for \eqref{mad}.  Therefore, when next we divide the relative entropy by $t$ and let $t\uparrow \infty$, we get
\begin{eqnarray}
0\leq \lim_{t\to\infty} \int P^V \log \frac{P^V}{P} &=& \lim_{t\to\infty} \frac 1{t}\int_0^t\id s \sum_{x,y}\rho_s(x) [k(x,y) - k_V(x,y)]\nonumber\\
 &=& \sum_{x,y} \nu^V(x) [k(x,y) - k_V(x,y)]\label{xx}
\end{eqnarray}
where $\nu^V$ is the stationary distribution of the process with rates $k_V$ (defined in \eqref{mad}). Since that expression is non-negative, the infimum of the last line over all possible potentials $V$ is zero.\\
  We can turn that around because there is a one-to-one relation between $V$ and $\nu^V$:  
Given a probability density $\rho$ we can find a potential $V_\rho$ such that the modified process $P^{V_\rho}$ with transition rates $k^\rho(x,y) := k_{V^\rho}(x,y)  $ has exactly that $\rho$ as its stationary distribution:
\[
\sum_y \left[ k^\rho(x,y)\,\rho(x)  - k^\rho(y,x)\,\rho(y) \right]=0, \quad\text{  for all } x
\]
Given $\rho$, such a potential $V_\rho$ is unique up to a constant; see \cite{mono}.\\

The previous considerations lead to the Donsker-Varadhan variational principle, \cite{DV}.
Consider the functional $\mathbb{D}(\rho)$ on probability distributions $\rho>0$,
\begin{equation}\label{dv}
\mathbb{D}(\rho) := \sum_{x,y} \rho(x) [k(x,y) - k^\rho(x,y)]
\end{equation}
which is a difference of expected escape rates as we had them around \eqref{esc}. Clearly, \eqref{dv} is a version of \eqref{xx}. 
Here comes the variational principle: the stationary distribution $\nu$ of the original process with rates $k(x,y)$ is the probability distribution that minimizes $\mathbb{D}(\rho)\geq 0$, or,
\[
\nu =\arg\min \mathbb{D}(\rho)
\]
In other words, the distribution minimizing the functional $\mathbb{D}$ defined in \eqref{dv} is the stationary distribution.  
Our point here is that the proposed variational principle involves minimizing an excess frenesy.  (We did not change the activation parameters to get the modified process.)  Or, the stationary distribution $\nu$ is the probability law for which a specific expected excess frenesy gets extremized.\\

  We close this section by noting that close-to-equilibrium the above (frenetic) variational principle becomes the minimum entropy production principle, \cite{minep}.  That goes as follow.\\
If the transition rates $k(x,y)$ are a smooth deformation of a finite irreducible system with rates satisfying the global detailed balance condition, then
\begin{equation}\label{mie}
\mathbb{D}(\rho) =  \frac 1{4}\left( \sigma^\epsilon(\rho) - \sigma^\epsilon(\nu)\right) + O(\epsilon^2\,d[\rho,\nu],d[\rho, \nu]^3) \geq 0
\end{equation}
where $\sigma^\epsilon$ is the entropy production rate functional.  For Markov jump processes, that is
\[
\sigma^\epsilon(\rho) = \sum_{x,y}\rho(x) k(x,y)\,\log\frac{\rho(x)k(x,y)}{\rho(y)k(y,x)}
\]
and the superscript $\epsilon$ indicates the nonequilibrium parameter in the transition rates $k(x,y)$ with respect to the reference equilibrium process.  The  $d[\rho,\nu]$ is the distance between $\rho$ and stationary distribution $\nu$. From \eqref{mie} we learn that, when close to equilibrium, minimizing $\mathbb{D}(\rho)$ is the same as minimizing the entropy production functional $\sigma^\epsilon(\rho) $ over $\rho$. We can see \eqref{mie} again as an instance of the merging of frenetic and entropic aspects close-to-equilibrium as discussed also around \eqref{norma}.\\   In a similar sense it has also been shown how in the case of detailed balance the functional \eqref{dv} gives an upper bound to the spectral gap, and hence informs about relaxation times in the approach to equilibrium.  See \cite{minep} for more details.

\subsection{Kinetic uncertainty and the Fisher metric}\label{kinu}
We start again with the calculation of  a relative entropy
\begin{equation}\label{reen}
s_\alpha := \int \id P_\alpha(\omega)\,\log \frac{\id P_\alpha}{\id P}(\omega)
\end{equation}
between a modified and an original dynamical ensemble over some time-interval $[0,t]$.  Here $P_{\alpha=0}=P$ where $\alpha$ is a small number parameterizing the modification. The expression \eqref{reen} is also commonly referred to as a Kullback-Leibler divergence between $P_\alpha$ and $P$ (here, dynamical ensembles).  We substitute the Taylor expansion of the action,
\[
A_\alpha := -\log \frac{\id P_\alpha}{\id P} =\alpha A'_0 + \frac{\alpha^2}{2}\,A''_0 + \ldots
\]
with the primes referring to differentiation with respect to $\alpha$ (at $\alpha=0$).  Therefore, from substituting in \eqref{reen}, 
\begin{eqnarray*}
s_\alpha &=& -\int \id P(\omega)\,\exp[-\alpha A'_0 - \frac{\alpha^2}{2}\,A''_0 + \ldots]\,[\alpha A'_0 + \frac{\alpha^2}{2}\,A''_0 + \ldots]\nonumber\\
&=& \alpha^2 \int \id P(\omega)\,\left(A'_0\right)^2 (\omega)+ \ldots
\end{eqnarray*}
where we have used that 
\[
\int \id P(\omega)\,\exp[-\alpha A'_0 - \frac{\alpha^2}{2}\,A''_0 + \ldots] = 1
\]
 For that same reason
\[
\int \id P(\omega)\,\left(A'_0\right)^2 (\omega) = \int \id P(\omega)\,A''_0 (\omega)\]
and hence the relative entropy \eqref{reen} equals
\begin{equation}\label{fin}
s_\alpha = \alpha^2\,\int \id P(\omega)\,A''_0 (\omega) = \alpha^2\,\langle A''_0\rangle
\end{equation}
to highest order in $\alpha\ll 1$.  The right-hand side of \eqref{fin} is the so called Fisher information metric (here in one dimension for simplicity, $\alpha\in \R$) and we see it is given by the second order in the action. In typical cases, such second order terms in the action are frenetic.\\
While frenesy thus clearly contributes in the Fisher metric, we emphasize that we deal here with distances in the spaces of dynamical ensembles.  Information geometry on the other hand usually considers thermodynamic landscapes with distances between static ensembles (usually Gibbs distributions) \cite{info}.  Here we move in a landscape which is formed by dynamical (kinetic) prescriptions.

Let us now take specific modifications, first by changing the activity parameters $a(x,y)\rightarrow (1+\alpha)a(x,y)$ in Markov jump processes; recall \eqref{apa}.  Clearly, only the frenesy in the action is affected, and only the activated traffic to second order in $\alpha$.  From \eqref{dd}, the (excess) activated traffic equals 
\[
\Delta \text{Act}(\omega) = \log(1+\alpha) \,{\cal T}(\omega) 
\]
and therefore $s_\alpha = \alpha^2 \langle {\cal T}\rangle$, the expected number of jumps in the considered time-window. If instead of having one fixed parameter $\alpha$ we would take $\alpha(x,y)$ depending on the bond, then we would get the Fisher metric related to the traffic over the different bonds.\\
Similarly, for underdamped diffusions we consider the modification
\[
\dot{v} = -\gamma(1+\alpha)v + F + \sqrt{2{\cal D}}\xi_t
\]
where the friction is increased by $1+\alpha$ without touching the diffusion constant ${\cal D}$.  Then,
the second order in the action is the kinetic energy: $s_\alpha = \alpha^2 \,\gamma^2\,\langle v^2\rangle/(4{\cal D})$.\\

The above was used by Di Terlizzi and Baiesi in the derivation of kinetic uncertainty relations, \cite{terl}.  The point is that the Fisher metric gives a general bound on the response of an observable when perturbing the dynamical ensemble from $P$ to $P_\alpha$: the Dechant-Sasa inequality gives almost directly that
\[
\left(\frac{\partial \langle O \rangle_\alpha}{\partial \alpha}\big|_{\alpha=0}\right)^2 \leq 2\,\text{Var}[O]\,\langle A''_0\rangle
\]
for an arbitrary path-observable $O=O(\omega)$ on $[0,t]$ with variance Var$[O]$, and with $\langle\cdot\rangle_\alpha$ the expectation in the perturbed process; see \cite{des,terl} for details. Observe that the (unperturbed) expectation $\langle A''_0\rangle$ is related to the frenesy as the above examples illustrate.  The result of \cite{terl} is therefore a kinetic uncertainty relation, in contrast with thermodynamic relations in e.g. \cite{udo2,hal2} which however appear to work best only close-to-equilibrium.  More recent discussions are contained in \cite{rold}.  The observation that dynamical activity controls bounds on fluctuations of counting observables such as traffic was formulated before in \cite{gar2}.  An uncertainty bound for underdamped Langevin dynamics in terms of both entropy production and dynamical activity was recently obtained in \cite{vu}.

\section{Response theory}\label{res}

Suppose we prepare the system of interest in a specific initial condition after which we apply a stimulus.
The goal of response theory is that one may in a systematic way predict or estimate and characterize physically the statistical response, preferably from observations in the initial condition.  Both the stimulus (or perturbation) and the observed quantity may be time-extensive or not.  Our approach is always the same, using dynamical ensembles, and is less analytic and more probabilistic compared with traditional approaches that are often following the formalities of quantum mechanics \cite{kub,green}.  The idea is that the stimulus changes the dynamical ensemble, so we modify the (original, unperturbed) reference probability Prob$_\text{ref}= P_0$ on paths to the perturbed one Prob $= P_\epsilon$ where $\epsilon$ abbreviates the spacetime amplitude of the perturbation.  We then have \eqref{dyen}, in the form
\[
\text{Prob}[\omega] = e^{-\Delta D(\omega) + \frac 1{2} \Delta S(\omega)}\,\text{Prob}_\text{ref}[\omega]
\]
where the excesses or differences indicated via ``$\Delta$'' are due to the perturbation over a time say $[0,t]$.  Per consequence we can compute averages in the perturbed ensemble via the formula
\begin{equation}\label{peens}
\langle O\rangle_\epsilon = \int P_0(\id\omega)\,O(\omega)\,e^{-{\cal A}(\omega)} = \langle O\, e^{-\Delta D + \frac 1{2} \Delta S}\rangle_0
\end{equation}
for some observable $O=O(\omega)$ depending on the path $\omega$ in the time $[0,t]$. The right-hand side is an average in the $\epsilon=0$ reference ensemble. It is assumed here that the initial conditions at time zero coincide for the perturbed and the original ensemble. We write the response as
\begin{equation}\label{peensb}
\langle O\rangle_\epsilon - \langle O\rangle_0=  \left<O\, \left[e^{-\Delta D + \frac 1{2} \Delta S} - 1\right]\right>_0
\end{equation}
with $\Delta D, \Delta S$ starting with order $\epsilon$.
The rest of the mathematics is to expand the exponential, e.g.  $\Delta D = \epsilon\,D'_0 + \frac{\epsilon^2}{2}D''_0 + \ldots,\quad \Delta S = \epsilon\,S'_0 + \frac{\epsilon^2}{2}S''_0 + \ldots$ as written in the simplest case for a single real parameter $\epsilon$. The equation \eqref{peensb} to second order in $\epsilon$ becomes,
\begin{eqnarray}\label{wonders}
\langle O\rangle_\epsilon - \langle O\rangle_0 &=& \epsilon\left\langle O \left[-D'_0 + \frac 1{2} S'_0 \right]\right\rangle_0 \\
&& +\;\frac{\epsilon^2}{2}\, \left\langle O \left[-D''_0 + \frac 1{2} S''_0  + (D'_0)^2 + \frac 1{4} (S'_0)^2 - D'_0S'_0\right]\right\rangle_0\nonumber
\end{eqnarray}
Similar expressions are readily obtained in the case of time-dependent perturbations as well. The real work is to simplify and to interpret the result. After all, the main purpose is to ``predict'' the response from ``measuring'' or ``estimating'' fluctuations in the reference or unperturbed system.  There, symmetries, and first of all, time-reversal (anti)symmetries play a major role.  We will see that in linear response around nonequilibria or in nonlinear response around equilibria, the frenesy contributes essentially.  For more detail, time-dependent extensions and discussions, also including examples and experimental verifications, we refer to \cite{fdr,njp,urna16,juan}.\\
 There are of course different versions of response theory.  We refer to \cite{falb,res1,res2,res3,res4,res5,res6,res7,res8,res9,res10,res11,res12,res13} for other approaches and many more results, but even then there are various other ideas.  In particular we mention the work of Komatsu and Nakagawa in \cite{naoko} for characterizing nonequilibrium stationary distributions, and the Harada-Sasa and Dechant-Sasa inequalities in \cite{hasa,dech}. The heart of response theory is not its formal appearance  -- in the end we are doing Taylor expansion assuming (and sometimes proving) convergence. Here, in this review on frenesy, we emphasize (only) the importance of the frenetic contribution in response.\\
  We start with the linear response theory around  nonequilibria.

\subsection{Linear response around nonequilibrium}
First order response can be read from \eqref{wonders} to be
\begin{equation}\label{linwon}
\langle O\rangle_\epsilon - \langle O\rangle_0 = \epsilon\left\langle O \left[-D'_0 + \frac 1{2} S'_0 \right]\right\rangle_0 
\end{equation}
As above, we write $D'_0,S'_0$ for the first derivatives evaluated at $\epsilon =0$. In the case where the observable is odd under time-reversal, $O\theta = -O$, and the reference process is time-reversal invariant, $\langle g(\theta\omega)\rangle_0 = \langle g(\omega)\rangle_0$ or $P_0(\omega) = P_0(\theta\omega)$, we have $\langle D'_0\, O \rangle_0 =0$ and only the entropic contribution survives in the linear response formula \eqref{linwon}.  That is then the origin of the fluctuation--dissipation relation for perturbations of reference equilibria, where we write with expectations $\langle\cdot\rangle_{\text{eq}} =\langle\cdot\rangle_\text{ref} = \langle\cdot\rangle_0$,
\begin{equation}\label{eqlin}
\langle O - O\theta\rangle_\epsilon =  \frac{\epsilon}{2}\,\left\langle [O - O\theta]\, S'_0\right\rangle_\text{eq} = \epsilon\,\left\langle O \,S'_0\right\rangle_\text{eq}
\end{equation}  
On the other hand, for perturbations around nonequilibrium, we need to include the frenetic contribution even in linear order.  For example taking as observable $O(\omega) = f(x_t)$ a function of the state $x_t$ at time $t$, we get
\begin{eqnarray}
\langle f(x_t) \rangle_\epsilon - \langle f(x_t) \rangle_0 &=& \frac{\epsilon}{2} \langle f(x_t)\,  S'_0(\omega) \rangle_0 - \epsilon\langle f(x_t)\,D'_0(\omega) \rangle_0\nonumber\\ 
&=& \epsilon\,\langle f(x_t)\,  S'_0(\omega) \rangle_0 - \epsilon\left< \,f(x_t)\,\left[ D'_0(\omega) + S'_0(\omega) /2\right]\,\right>_0\label{kuku}
\end{eqnarray}
where the last line rewrites the linear response formula so that the first term on the right-hand side contains the correlation as in  the standard Kubo formula (\cite{kub,njp} for linear response around equilibrium).  Observe that via time-reversal invariance in equilibrium, $\langle \,f(x_t)\,\left[ D'_0(\omega) + S'_0(\omega) /2\right]\,\rangle_\text{eq} = \langle \,f(x_0)\,\left[ D'_0(\omega) - S'_0(\omega) /2\right]\,\rangle_\text{eq} = 0$ because of \eqref{norma}.\\
To understand why the first term on the right-hand side of \eqref{kuku} is Kubo-like, it suffices to take some physical examples.  Suppose for example that the perturbation is of the potential form as in \eqref{frda} with $F(x) = g(x) + \nabla V(x)$; then the excess entropy flux per $k_B$ is $h\beta(V(x_t)-V(x_0))$ and when $h=h_s$ is a time-dependent vector we find
\[
\langle f(x_t)\,  S'_0(\omega) \rangle_0 =\beta\int_0^t\id s \,h_s\cdot\,\langle f(x_t)\,\nabla V(x_s) \rangle_0
\]
as is the Kubo-form. Note the prefactor $\beta$ corresponding to the inverse temperature of the surrounding thermal bath.  While the correction to the linear response is additive compared to the equilibrium expression, we can of course albeit artificially turn it into a multiplicative correction by writing \eqref{kuku} as
\[ 
\langle f(x_t) \rangle_\epsilon - \langle f(x_t) \rangle_0  =\epsilon\,\beta\left(1  -  \frac{\langle \,f(x_t)\,\left[ D'_0(\omega) + S'_0(\omega) /2\right]\,\rangle_0}{\langle f(x_t)\,  S'_0(\omega) \rangle_0}\right)\,\int_0^t\id s \,h_s\cdot\,\langle f(x_t)\,\nabla V(x_s) \rangle_0 \label{kukul}
\] 
Then, the prefactor $\beta\,(\cdot)$ may be called an {\it effective} inverse temperature.  That is how effective temperatures have appeared often in the literature, of course mostly depending on the observable $f$; see e.g. \cite{cug}.  For example, if $\langle \,f(x_t)\,D'_0(\omega)\rangle_0 \simeq 0$ then the effective temperature $T_\text{eff} \simeq 2T$ is double the thermodynamic surrounding temperature. Note also that this strategy of defining effective temperatures differs from expressions like in \eqref{tef} for characterizing the population inversion.  Here we learn how they abbreviate drastically the frenetic contribution.\\

Let us take the case of a Markov jump process, as in Section \ref{timesym}.\\
Note that a perturbation (here time-independent)
\begin{equation}\label{per}
k(x,y) \rightarrow k_\ep(x,y) = k(x,y)\,[1 + \epsilon a_1(x,y) + \frac{\epsilon}{2}\,s_1(x,y)]
\end{equation}
with symmetric $a_1(x,y)=a_1(y,x)$ and antisymmetric $s_1(x,y)=-s_1(y,x)$,
is equivalent to the changes 
\begin{equation}\label{pra}
s(x,y)\rightarrow s(x,y) + \epsilon\,s_1(x,y),\quad  a(x,y)\rightarrow a(x,y) + \epsilon\,a_1(x,y)
\end{equation}
to linear order in $\epsilon$.  Then, always to linear order in $\epsilon$, the excess frenesy equals
\begin{equation}\label{edal}
D(\omega) = -\epsilon\sum_s a_1(x_{s^-},x_s) + \epsilon\int_0^t \id s \sum_y k(x_s,y)[a_1(x_s,y) + \frac 1{2} s_1(x_s,y)]
\end{equation}
which is responsible  for the frenetic contribution to linear response.  For the expectation to linear order in $\epsilon$ we get,
\begin{eqnarray}\label{linre}
\langle O\rangle_\epsilon - \langle O\rangle_0 &=& \frac{\epsilon}{2}\, \langle \sum_s s_1(x_{s^-},x_s)\,O(\omega)\rangle_0\\
&+& \epsilon\;\langle\left[\sum_s a_1(x_{s^-},x_s) - \int_0^t \id s \sum_y k(x_s,y)[a_1(x_s,y) + \frac 1{2} s_1(x_s,y)]\right]\,O(\omega)\rangle_0 \nonumber
\end{eqnarray}
for an arbitrary path-observable $O$ over time $[0,t]$.\\
Applications and more interpretations follow in Section \ref{appil}.

\subsection{Nonlinear response around equilibrium}\label{2or}

There is no problem to continue and to expand to higher order.   Let us take the reference condition to be the one of equilibrium with expectations $\langle\cdot\rangle_\text{eq}$, and let us suppose that $ S_\epsilon = \epsilon \,S'_0$. That assumption is natural and generally valid when adding external fields or potentials as perturbations.  Then, from \eqref{wonders}, as $S_0''=0$ and both $D_0''$ and $(S_0')^2$ are even under time-reversal,
\begin{equation}\label{2n4}
\langle O - O\theta\rangle_\ep = \epsilon \,\langle S'_0(\omega)\, O(\omega)
\rangle_{\text{eq}} - \epsilon^2\,\langle D'_0(\omega)\,S'_0(\omega)\, O(\omega)\rangle_{\text{eq}}
\end{equation}
For a state observable, that is when $O$ only depends on the state at a single time, $O(\omega)=f(x_t)$, we have $O\theta(\omega)= O(\pi x_0)$ (where the extra involution $\pi$ could be the flipping of all velocities as encountered in inertial or underdamped dynamics), and we can use $\langle f(\pi x_0)\rangle_{\text{eq}} = \langle f(x_0)\rangle_{\text{eq}} = \langle f(x_t)\rangle_{\text{eq}}$.  Applying formula \eqref{2n4} to that case we obtain the extension  to second order of the traditional Kubo formula,
\begin{equation}\label{kubo2}
\langle f(x_t)\rangle_\ep - \langle f(x_t)\rangle_{\text{eq}} =  \varepsilon\,\langle S'_0(\omega)\,f(x_t)
\rangle_{\text{eq}} - \varepsilon^2\,\langle D'_0(\omega) \,S'_0(\omega)\, f(x_t)\rangle_{\text{eq}}
\end{equation}
That equation \eqref{kubo2} holds for general time-dependent
perturbation protocols as well; see \cite{pccp}.\\
Similarly, when $O(\theta\omega) = -O(\omega)$  for time-integrated particle or energy currents $O(\omega) = J(\omega)$, we get from \eqref{2n4} the extended Green--Kubo formula, \cite{njp},
\begin{equation}\label{gk}
\langle J\rangle_\ep = \frac{\varepsilon}{2} \,\langle S'_0(\omega)\, J(\omega)
\rangle_{\text{eq}} - \frac{\varepsilon^2}{2}\,\left < D'_0(\omega)\,S'_0(\omega)\, J(\omega)\right >_{\text{eq}}
\end{equation}
We can also take $O(\omega) = S'_0(\omega)$ in \eqref{2n4} and
\[
\langle S'_0 \rangle_\ep = \frac{\varepsilon}{2} \,\left < (S'_0)^2
\right >_{\text{eq}} - \frac{\varepsilon^2}{2}\,\left < D'_0\,(S'_0)^2\right >_{\text{eq}}
\]
where we see that the sign of the second-order term depends on an entropy--frenesy correlation in equilibrium, correcting the fluctuation--dissipation relation (which only has the first term on the right-hand side).
From second order onward, frenesy plays its role in response. Mutilated ensembles that are produced via maximum entropy principles on path-space and do not take the frenesy as an observable path-observable will fail at this point.   We agree with \cite{cpr}  that maximum entropy methods on spacetime trajectories (the 1980 maximum caliber principle as formulated by E.T.~Jaynes) remain a viable principle when taking into account sufficient variables and constraints (on them).
 
 \subsection{Applications}\label{appil}

Response theory has obviously plenty of applications.  We select some of those with a more theoretical or conceptual character.

\subsubsection{Modified Einstein relations}\label{mer}

There are a number of so called Einstein relations in statistical physics.  The two that we discuss below have to do with response.  The terminology is somewhat confused here.  To avoid misunderstandings we speak about (1) the Sutherland-Einstein relation when meaning the linear response for mobility, and (2) the Einstein relation when dealing with the connection between friction and noise.  Both need to be extended to take into account the frenetic contributions when confronted with the nonequilibrium world.\\
  The first meaning of Einstein relation discussed below is between diffusion and mobility.  That so called Sutherland-Einstein relation is a direct application of linear response theory and together with the (also related) Johnson-Nyquist relation was among the first examples of the so called (first)  fluctuation--dissipation relation.\\
  The second Einstein relation to be discussed concerns the relation between the friction and the noise amplitudes for Brownian particles.  Experience tells us that a probe (e.g. a colloid or other mesoscopic object) suspended in and moving through an environment of many much faster and smaller particles experiences both friction and statistical fluctuations.  A systematic derivation of that Einstein relation requires a suitable scale of description as for example in the Van Hove scheme of a weak coupling limit, \cite{vanh}.  Then, when the environment is a thermal equilibrium bath, noise and friction are not independent but connect in what is called the second fluctuation--dissipation relation, or indeed the Einstein relation.\\

The first fluctuation--dissipation relation centers on the relation between mobility and diffusion. The question is how transport coefficients can be guessed from the fluctuation properties in the unperturbed system.  In equilibrium the transport coefficients can be expressed as Helfand moments that are mean square deviations. That is, the Green-Kubo relation gives so called Einstein-Kubo-Helfand expressions for transport coefficients, which can be seen as (generalized) diffusion constants, \cite{hel,gasp}.  In the case of particle diffusion and its relation with mobility (from measuring the induced velocity after applying a driving external field), we speak about the Sutherland--Einstein relation.  It holds generally around equilibrium.  Around nonequilibrium (or starting in second order around equilibrium) that equivalence between mobility and diffusion is violated, and the Sutherland--Einstein relation must be corrected with a frenetic contribution.  A larger exploration of the issues is contained in \cite{proc,soghra,gal}.\\
 More explicitly, we take a particle of mass $m$, which diffuses in a heat bath according to the Langevin dynamics for the position $\vec{r}_t$  and the velocity $\vec{v}_t$,
 \begin{eqnarray}
 \dot{\vec{r}}_t  &=& \vec{v}_t \label{general}\\
  m\dot{\vec{v}}_t &=& \vec{F}(\vec{r}_t,\vec{v}_t)-\gamma m \vec{v}_t + \sqrt{2m\gamma T}\,\vec{\xi}_t \nonumber
 \end{eqnarray}
The external force $\vec{F}$ depends periodically on the position $\vec{r}$, and there is no confining potential (which is relevant for the purpose of this discussion about transport). The vector $\vec{\xi}_t$ is standard Gaussian white noise,
 with mean zero $\left<\xi_{t,i}\right> = 0$ and covariance $\left<\xi_{t,i}\xi_{s,j}\right> = \delta_{i,j}\delta(t-s)$.\\

 The diffusion (matrix) ${\cal D}(t)$ is defined as
 \[
  {\cal D}_{ij}(t) = \frac{1}{2t}\Big<(\vec{r}_t-\vec{r}_0)_i;(\vec{r}_t-\vec{r}_0)_j\Big>
   \]
with the subscripts for the components of the corresponding vectors. We continue to use the notation that for observables $A$ and $B$,   $\Big<A;B\Big>=\Big<AB\Big> -\Big<A\Big>\Big< B\Big>$.
We expect a large time limit as the diffusion matrix
 \[
  {\cal D}_{ij} = \lim_{t\to\infty}{\cal D}_{ij}(t) 
 \]
 On the other hand there is the mobility (matrix) function $M(t)$ which is obtained by adding to the dynamics (\ref{general}) a constant (but small) force $\vec{f}$, replacing $\vec{F}(\vec{r},\vec{v})
 \rightarrow \vec{F}(\vec{r},\vec{v}) + \vec{f}$. The mobility
 is the change in the expected displacement:
 \[ M_{ij}(t) = \frac{1}{t}\left.\frac{\partial}{\partial f_j}\Big<({\vec r}_t-{\vec r}_0)_i\Big>^{f}\right|_{\vec{f}=0} \]
 where $\langle\cdot\rangle^f$ is the average in the dynamics with the extra force $\vec{f}$.
 Again in the large-time limit, we get the mobility
 \[ M_{ij} = \lim_{t\to\infty}M_{ij}(t) \]
as the linear change in the stationary velocity by the addition of a
 small constant force.\\
The Sutherland--Einstein relation tells that diffusion matrix and mobility are proportional,
\begin{equation}\label{einstein} 
M_{ij} = \frac{1}{T}{\cal D}_{ij} 
\end{equation}
which is however only to be expected near equilibrium and there it is a consequence of the usual linear response theory as in \eqref{linwon}. When the system is not in equilibrium, frenetic terms show up so that the mobility and diffusion constants are no longer proportional.  It was derived in \cite{proc} that the nonequilibrium modification of the Sutherland-Einstein relation is given by
\begin{eqnarray}
 && M_{ij} = \frac{1}{T}{\cal D}_{ij} - \lim_{t\to\infty}\frac{1}{2\gamma mT }\,\int_0^t \id s\,\Big<\frac{({\vec r}_t - {\vec r}_0)_i}{t};F_j(\vec{r}_s,\vec{v}_s)\Big>\label{genresult5}
\end{eqnarray}
The correction to the equilibrium mobility--diffusion relation is frenetic and is measured by a spacetime correlation between applied forcing and displacement.  That is the last expectation in the right-hand side of \eqref{genresult5}. The deviation with respect to the Sutherland-Einstein relation is second order in the nonequilibrium driving.  See also the discussion in \cite{sar}.\\
 
So far we have discussed the response on statistical expectations after giving a stimulus in a specific reference condition.  As an example of a stimulus we may consider the motion of a probe in the system. The system plays the role of the medium now with the motion of the probe perturbing the system.  It is of course a time-dependent perturbation.  The system responds and that feeds back to the probe motion, making friction and noise.   That scenario has been recently investigated in a number of papers starting from nonequilibrium response, \cite{jsp,stefan,leipzig,krueger}.\\
We consider a process $x_t$ representing the time-dependent degrees of freedom of a bath in which there moves a probe with position $Y_t$. The bath typically has many degrees of freedom $x_t(i), i=1,\ldots,N$ for large $N$ and the coupling with the probe is weak.  We want to understand what of the bath enters the dynamics of the probe when integrating out the bath degrees of freedom. We fix time $t$ and look at times $s\leq t$ before time $t$ and starting some time in the past.  The dynamical ensemble for the (bath) $x-$trajectories $\omega$ is given by our main relation \eqref{dyen},
\begin{equation}\label{app}
P(\omega|Y_s, s\leq t) = \exp [-D(\omega) + \frac 1{2} S(\omega)]  \,  P(\omega|Y_s=Y_t, \text{ for all } s\leq t) 
\end{equation}
where the left-hand side is conditioned on a(n arbitrary) probe trajectory $(Y_s)^t$, while the probability in the right hand-side is the reference probability on bath trajectories supposing the probe has always been at the (final) position $Y_t$.  Clearly, the frenesy $D$ and the entropy flux $S$ are (as always) excess quantities and they depend also on the probe trajectory (not indicated).  We need to expand them (here only) to first order around the reference probe trajectory $Y_s\equiv Y_t$.  We assume for simplicity that the probe position only enters via an interaction potential $U(Y,x) = 
\sum_{i=1}^N u(Y-x(i))$ which, in other words, constitutes the only coupling between bath and probe.  That interaction is small in the appropriate sense and we also assume that the bath is overdamped and diluted, so we can treat the $x_t(i)$ as independent from each other.  Then we can use the expressions around \eqref{overd}--\eqref{fred} with bath-dynamics, 
\[
\dot x_s = \chi\, F(x_s,Y_s) + \sqrt{2\chi k_BT}\,\xi_s,\qquad s\leq t
\]
from now on in one-dimensional notation and per bath-particle.  Here.
\[
 F(x_s,Y_s) = u'(Y_s-x_s) + f(x_s),\quad u'(Y_s-x_s) = u'(Y_t-x_s) + (Y_s-Y_t)\,u''(Y_t-x_s)  
 \]  
 where we already made the expansion to linear order around the probe trajectory $Y_s\equiv Y_t, s\leq t$.  The force $f$ on each bath degree of freedom may be nonconservative.  The bath is indeed itself here an open system which is driven out-of-equilibrium and dissipates in a thermal equilibrium environment represented in the mobility $\chi$, by its temperature $T$ and through the standard white noise $\xi_s$.
 We thus have in \eqref{app}, following \eqref{frda},
\begin{eqnarray}\label{sda}
S(\omega, (Y_s)^t) &=&  \beta\int^t \id x_s \,(Y_s-Y_t)\,\circ u''(Y_t-x_s),\\
 D(\omega,(Y_s)^t) &=& \beta\chi\int^t\id s \,(Y_s-Y_t)\,u''(Y_t-x_s)\, [u'(Y_t-x_s) + f(x_s)] \nonumber\\
 -&& \chi\int^t\id s\,(Y_t-Y_s)u''' (Y_t-x_s) \label{dsa}
\end{eqnarray}
which are the entropy flux caused by the motion of the probe and the first order (in $Y_s-Y_t$) term in (excess) frenesy as function of the bath-trajectory $\omega$. One needs to sum over all the bath-particles.  Interactions between the bath-particles do not matter if not influenced directly by the presence of the probe.\\
On the other hand, the force of each bath-particle on the probe is
\begin{eqnarray}\label{fobi}
-u'(Y_t-x_t) &=&  - \int\id \omega \,
P(\omega|Y_s, s\leq t)\,u'(Y_t - \omega_t) + \zeta_t,\nonumber\\
 \zeta_t =\zeta_t((Y_s)^t,x_t) &:=& \int\id \omega \,
P(\omega|Y_s, s\leq t)\,u'(Y_t - \omega_t) - u'(Y_t- x_t)  
\end{eqnarray}
in which we introduced a fluctuation term $\zeta_t$  which still depends on the bath-variables $x_t$ but has mean zero for every probe trajectory $(Y_s)^t$.  For $P(\omega|Y_s, s\leq t)$ we are ready to use \eqref{app}, and we get the force on the probe in the form
\begin{eqnarray}\label{fob}
-u'(Y_t-x_t) &=&  - \langle u'(Y_t-\omega_t)\rangle^{Y_t} -\left\langle S(\omega, (Y_s)^t)\,;\,u'(Y_t-\omega_t)\right\rangle^{Y_t}+ \zeta_t\\ 
&& +\left\langle D(\omega,(Y_s)^t)+\frac 1{2}\,S(\omega, (Y_s)^t)\,;\,u'(Y_t -\omega_t)\right\rangle^{Y_t}\nonumber
\end{eqnarray}
where the average $\langle\cdot\rangle^{Y_t}$ is taken over the stationary bath-particles with the probe at rest in $Y_t$.  We use the notation $\langle A\,;B\rangle =\langle AB\rangle - \langle A\rangle\langle B\rangle$ to denote the covariance which is allowed because $\frac1{2}\langle S(\omega, (Y_s)^t) \rangle^{Y_t} = \langle
D(\omega,(Y_s)^t)]\rangle^{Y_t}$ as always.
The first term on the right-hand side of \eqref{fob}  is the zero order force 
\[
\langle u'(Y_t-\omega_t)\rangle^{Y_t} =  \int\id \omega \,
P(\omega|Y_s=Y_t, s\leq t)\,u'(Y_t-\omega_t) 
\]
and corresponds to the systematic force obtained under infinite time-scale separation between bath and probe.  It is the statistical force on the probe which will be discussed more in Section \ref{statf}.\\
In the next (first) order, inside the entropic contribution, 
we have
\[
\int^t (Y_s-Y_t)\,\id x_s \,\circ u''(Y_t-x_s)= -\int^t \id s (Y_s-Y_t) \frac{\id}{\id s} u' (Y_t- x_s) 
\]
so that
\[
\left\langle S(\omega, (Y_s)^t)\,;\,u'(Y_t-\omega_t)\right\rangle^{Y_t} = \beta \int^t \id s \,\dot{Y}_s \,\left\langle u' (Y_t - \omega_s)\,;\, u'(Y_t-\omega_t)\right\rangle^{Y_t} 
\]
We see that the entropic contribution in the first line of \eqref{fob} amounts to introducing a friction term with memory kernel given by the force-force time-correlation function.  
The last term in the first line of  \eqref{fob} is the  noise introduced in \eqref{fobi} and given in zero order as
\[
\zeta_t^0(Y_t) = \langle u'(Y_t-\omega_t)\rangle^{Y_t} -u'(Y_t-x_t), \quad  \langle \zeta_t^0(Y_t) \rangle^{Y_t}=0
\]
while the time-correlations are
\[
\langle \zeta_t^0(Y_t)\zeta_s^0(Y_t) \rangle^{Y_t} = \langle u' (Y_t-x_s)\, ;\, u'(Y_t-x_t)\rangle^{Y_t} 
\]
As a summary, the induced force on the probe at time $t$ is
\begin{eqnarray}
-\langle u'(Y_t- x)\rangle^{Y_t} &&- \beta \int^t \id s \,\dot{Y}_s \,\left\langle u' (Y_t - \omega_s)\,;\, u'(Y_t-\omega_t)\right\rangle^{Y_t} +  \zeta_t^0(Y_t)\nonumber\\
+&&\left\langle D(\omega,(Y_s)^t)+\frac 1{2}\,S(\omega, (Y_s)^t)\,;\,u'(Y_t -\omega_t)\right\rangle^{Y_t}\label{frco}
\end{eqnarray}
We conclude therefore that it is the entropic term that produces the Einstein relation between the noise kernel and the friction memory (first line in \eqref{frco}).
For reversible dynamics where the second line in \eqref{frco} is absent there follows the well-known result that the friction matrix equals the force covariance, evaluated in equilibrium with the particle at rest at $Y_t$.
At the same time, we see how it is exactly the frenetic contribution that breaks the Einstein relation.  The modified Einstein relation can of course be reconstructed from the previous formul{\ae}.  The second line in \eqref{frco} contains the frenetic contribution and it will change the friction, such as to give e.g. negative contributions.\\

\subsubsection{Resolving kinetic differences}\label{resk}

Consider a box with $N(t)$ particles at time $t$. We think of a gas in a box which is open at its boundary for diffusion in equilibrium at temperature $T$ and chemical potential $\mu$. The system is therefore in thermal and chemical equilibrium with fixed volume, chemical potential and temperature.  Suppose that at time zero we increase the chemical potential, say from $\mu$ to $\mu + \delta$ at fixed $T$.  The particle system will relax to the new equilibrium, and the expected particle number will follow: $\langle N(t)\rangle$ will change in time.  We can describe the response in the linear regime and get
\[
\langle N(t) \rangle - \langle N\rangle_\text{eq} = \beta\delta\,\int_0^t \id s\,\big< J(s);N(t)\big>_\text{eq} 
\]
where $J(s)$ is the particle current at time $s$ into the environment, and $\langle\,\cdot\,\rangle_{\text{eq}}$ is the expectation in the original equilibrium process.  In fact, we can still rewrite that by using that $\int_0^t\id s J(s) = N(t) - N(0)$,
\[
\langle N(t) \rangle - \langle N\rangle_\text{eq} = \frac{\beta\delta}{2}\,\left< [N(t)-N(0)]^2\right>_\text{eq} 
\]
which is recognized as the fluctuation--dissipation relation applicable for small $\delta$.  Note that there is no need to specify the type of interaction or the types of particles; the relation only depends on the original chemical potential and its increase.  In that sense there is no difference whether the gas of bath particles smells of champagne or of anything else.\\
That stops being true in second order around equilibrium where the frenetic contribution enters, as discussed under Section \ref{2or}.  It now matters what are the exit and entrance rates of the particles, which involves kinetic information beyond the change in (thermodynamic) chemical potential.  In other words, there may be different {\it kinetic} ways to increase the bath chemical potential and they will give a difference for the time-dependence of  $\langle N(t) \rangle - \langle N\rangle_\text{eq}$ in second order around equilibrium ($\delta^2$).  Then, as e.g explored in \cite{pccp}, the total number of particle exchanges between the
system and the reservoir enters, which is the time-symmetric traffic.  Let us do the calculation for an open symmetric exclusion process.\\

In the exclusion process we consider a lattice interval $\{1,\ldots,N\}$ where each site $i$ can be occupied by at most one particle; $\eta(i)\in \{0,1\}$.  The particles can jump to their left or right nearest neighbor site provided no particle is there.  At the two boundary sites, particles can enter or exit.  We take the rates of a particle entering to left and to right edges, respectively as $\alpha \,(1-\eta(0))$ and $\gamma \,(1-\eta(N))$ for parameters $\alpha,\gamma >0$, and the rates of exiting from left and from right are then 
\[
\alpha\, e^{-\beta\mu}\, \eta(0),\qquad \gamma \,e^{\-\beta\mu}\,\eta(N)
\]
respectively, fixing thereby the environment as one equilibrium chemical reservoir at inverse temperature $\beta$ and chemical potential $\mu$. Note that $\alpha$ and $\gamma$ are still allowed to depend on $\mu$ or $\beta$. No matter how, the stationary condition is equilibrium with a product distribution as stationary reversible measure with density $(1+ \exp(\beta\mu))^{-1}$  (independent of $\alpha,\gamma$).  Starting from there at time zero, we let the dynamics run at a slightly different chemical potential $\mu\rightarrow \mu+\delta$.    In second order in $\delta$ we need to take into account the excess frenesy.  We assume first that $(\alpha,\gamma)$ do not depend on the applied chemical potential.   The excess in escape rates is then
\[
\Delta \xi(\eta) = \left(e^{-\beta(\mu+\delta)}-e^{-\beta\mu}\right)\,[\alpha\, \,\eta(0) + \gamma\, \eta(N)]
\]
while the excess in activated traffic becomes
\[
\Delta\text{Act} = \left(e^{-\beta(\mu+\delta)/2}-e^{-\beta\mu/2}\right)\,(\alpha \,I_0 + \gamma\,I_N)
\]
where $I_0$ and $I_N$ are the number of exchanges (time-symmetric traffic) at the left, respectively the right boundary.  Clearly those excesses would be different when $\alpha$ or $\gamma$ would depend on the chemical potential as well.  In fact, the variation with $\mu$ in $(\alpha,\gamma)$ would appear explicitly in $\Delta \xi$ and $\Delta $Act. Hence, the second order around equilibrium will depend on the specific kinetics through the frenetic contribution. The excess entropy flux on the other hand is purely thermodynamic as the product of the change in chemical potential times the net current of particles into the system.  The frenesy summarizes the influence of the detailed kinetics and delivers the correct response, here in second order around equilibrium.  Frenesy allows to sense the difference between champagne and water, which appears to be important.

\subsubsection{Small and negative susceptibilities}\label{smalln}

The Helfand form of the Green-Kubo relations, \cite{hel}, giving the (particle, energy or momentum) currents in linear order around equilibrium explicitly shows the positivity of the conductivity matrix.  After all, the Onsager matrix of linear response coefficients must be positive in accordance with the second law, \cite{dGM}.  Yet, when dealing with nonlinear response or with linear response around nonequilibria, there enters the frenetic contribution to the response.  In all it may certainly cause non-monotone behavior of the currents as a function of external field.  There are then regimes of very small and of negative differential conductivity. We had an elementary scenario in Example \ref{rw}. That scenario is in fact even wider, including also specific heats, compressibilities or chemical reactivities.  It also includes aspects of so called resilience in ecological systems, \cite{hol}.\\

The example \eqref{rw} already showed how trapping may lead to negative differential conductivities.  There are by now a large number of examples, including responses to temperature and chemical affinities; see \cite{zia,gar,negheatcap,oliver,chemfalasco,hao}.  For example, in \cite{hao} one sees modifier activation--inhibition switching in enzyme kinetics. Interestingly, one can be specific and quantitative about those non-monotonicities by showing the frenetic contribution explicitly and how it counteracts the entropic part in the linear response. The general structure is of course always the same, as we take it from \eqref{wonders} in linear order ($\epsilon$),
\begin{equation}\label{linwone}
\langle O\rangle_\epsilon - \langle O\rangle_0 = \epsilon\left\langle O \,\left[-D'_0 + \frac 1{2} S'_0 \right]\right\rangle_0 
\end{equation}
Taking for example as observable $O=S'_0$ (which is typically proportional to some current), we get in linear order,
\[
\langle S'_0\rangle_\epsilon - \langle S'_0\rangle_0 = \frac{\epsilon}{2} \langle (S'_0)^2\rangle_0 - \epsilon \langle S'_0\,D'_0\rangle_0 
\]
If therefore there is a positive correlation between the linear excesses in entropy flux and in frenesy in the {\it original} dynamics, the frenetic contribution adds negatively.  The amount of that negative frenetic contribution differs with the distance from equilibrium.  If $\langle \cdot\rangle_0$ refers to a time-symmetric process, as in steady equilibrium, then 
$\langle S'_0\,D'_0\rangle_0 =0$ by the time-antisymmetry of the product $S'_0\,D'_0$.  That is also why in and close-to-equilibrium, $\langle S'_0\rangle_\epsilon - \langle S'_0\rangle_0\geq 0$ always.  There are thus two conditions for getting a negative susceptibility for the observable $S'_0$: (1) one needs to be sufficiently away from equilibrium, and (2) one needs a positive correlation 
$\langle S'_0\,D'_0\rangle_0 >0$ in the original process.\\
If the perturbation is not affecting the activity parameters (or noise amplitudes), the frenesy only picks up the escape rates (see \eqref{fre}), and we typically need a positive correlation between the excess frenesy and the excess entropy flux.  If the dynamical ensemble away from equilibrium did not take into account the frenesy (but only had the entropy fluxes in the action), then negative response in the entropy flux would be impossible. By now, the above scenario has been observed and studied in a great number of cases; see e.g. \cite{gar,oliver,sar,chemfalasco,negheatcap}.  That is not to say that (absolute) negative conductivities (around equilibrium) would be impossible, see e.g. \cite{muk};  there, the entropy flux incorporates different currents and some may respond negatively to pushes.

We see also from \eqref{linwone}, as in Section \ref{resk}, that perturbations which are thermodynamically equivalent (same $S'_0$) still obtain a different response thanks to the frenetic contribution (different $D'_0$).
Sensing is thus improved via that kinetic effect.  
On the other hand, homeostasis or the condition of very weak susceptibility of certain observables can be characterized as the near orthogonality of the observable $O$ and the excess action,
$O \perp  \left[-D'_0 + \frac 1{2} S'_0 \right]$, in the sense of a vanishing right-hand side in \eqref{linwone}.  It means that the original (unperturbed) correlation between the observable and the excess frenesy is about half its correlation with the excess entropy flux.  Of course, such points of zero susceptibility are also reached when moving from a regime of positive to negative susceptibility (right-hand side of \eqref{linwone}).\\

\subsubsection{Statistical forces}\label{statf}

As a final application of response theory we look at the stationary distributions themselves, and how they change under perturbations.  That can be developed systematically, \cite{col}, but here we look at the linear response as applied for the problem of characterizing statistical forces.\\
We come back to the physics around Example \ref{crex}, where a statistical force is computed and it is asked whether that force generates a probe-current.
The terminology about so called statistical forces is not quite fixed. We do not have in mind here fluctuating forces as have mean zero under averaging.  Nor do we necessarily mean entropic or thermodynamic forces, because those are near--equilibrium quantities.  But we do mean systematic forces induced by a bath or medium on a probe or slower variable.  The latter refers to the infinite time-scale separation which is involved, and as we had it already in the expansion of Section \ref{mer}.  We will always work in that quasistatic regime here, and no friction or noise will be considered (see indeed Section \ref{mer} for that).\\

We start from a mechanical force arising from an interaction potential on some level of description.  Keeping the notation of Section \ref{mer} we have a huge number of bath variables $x = x(i), i=1,\ldots, N$, interacting with `probes' $Y_\alpha, \alpha=1,\ldots,n$ (possibly a smaller number $n\ll N$ of different more macroscopic particles).  Thinking for simplicity of all these variables as one-dimensional positions, we introduce a bath-probe coupling via interaction potential, 
\begin{equation}\label{pote}
U_\lambda(x,Y_\alpha) = U_0(x) + \lambda\sum_{i=1}^N u_{i\alpha}(x(i)-Y_\alpha)
\end{equation}
 where $\lambda$ denotes the coupling strength. In what follows we assume for simplicity that the probes are identical and that the bath-particles are also identical, so that the two-body interaction $u$ does not depend on $(i,\alpha)$.  The bath is supposed to be much faster, reaching stationarity with density $\rho_Y=\rho^\lambda_Y$ much before the probe(s) move significantly.  As a result there is a statistical (or systematic or mean) force $f=f^\lambda$ on the probe(s): when the latter occupy positions $Y$, the $\alpha-$th probe experiences  (always in one-dimensional notation),
\begin{equation}\label{staf}
f_\alpha(Y) = \lambda\,\sum_{i}\int \id x \,\rho_Y(x) \,u'(x(i)-Y_\alpha)
\end{equation}
which is the average force over the stationary density in the nonequilibrium medium or bath.  We will not use a dynamics to characterize that density $\rho_Y$ and it would be near to impossible to calculate it anyway, especially for interacting systems.  The big exception is equilibrium.
It is easy to calculate that, when $\rho_Y(x) \sim \exp[-\beta U(x,Y)]$, then $f(Y) = -\nabla_Y{\cal F}(Y)$ for free energy ${\cal F}(x) = -k_BT\log\int \id x \exp[-\beta U(x,Y)]$.  That is the equilibrium case, where the statistical force is derivable from a thermodynamic potential.\\

When the bath is under nonequilibrium driving, then the stationary distribution is very different possibly from the Gibbsian prescription with interaction potential $U$ and thermodynamic temperature $T$.  In nonequilibrium, stationary distributions may pick up detailed kinetic information as we have seen for example in Section \ref{pops}.  As a consequence, many such features as the gradient character of the statistical force \eqref{staf}, its additivity over various baths or its locality are no longer (and most often are not) guaranteed. Interestingly, those differences arise by the very presence  and the nature of frenetic contributions, as we now briefly indicate.  See \cite{nong,stf} for more discussion.\\

In order to characterize \eqref{staf} we need to bring the stationary density in a form which is physically meaningful, and which enables to connect features of the statistical force with the essential aspects of the bath dynamics. Here we concentrate on the dependence of $\rho^\lambda_Y$ on probe positions $Y$ and coupling constant $\lambda$ as can be obtained from linear response theory around nonequilibrium and is the subject of \cite{nong}.  The key is again the set-up via dynamical ensembles and the formula \eqref{dyen}.\\

 We always assume that the nonequilibrium medium enjoys a unique density with exponentially fast relaxation from all possible initial conditions.  The essence is to get a go on the response of the density $\rho_Y=\rho_Y^\lambda$ when changing $(Y,\lambda)$. We need to work with two ensembles, one corresponding to $(Y^{(1)},\lambda_1)$ with expectations $\langle \cdot \rangle_1$, and the modified ensemble corresponding to $(Y^{(2)},\lambda_2)$ with expectations $\langle \cdot \rangle_2$.  The corresponding densities in the bath at positions $x$ at time $t$ can be related by using
\begin{equation}\label{eg}
p^t_{2}(x) = \left< \delta(x_t-x)\right>_2 = \int e^{-{\cal A}(\omega)}\, \delta(x_t-x)\,\id P_1(\omega)
\end{equation}
where the action ${\cal A}=D_2-D_1 + (S_1-S_2)/2$ connects the two ensembles.  When we
divide by $\langle \delta(x_t-x)\rangle_1$ we get for all times $t$, even very large $t$ compared to the relaxation time, that
\begin{eqnarray}\label{ppt}
\frac{p^t_{2}(x)}{p^t_{1}(x)} &=& \langle e^{-{\cal A}(\omega)}\, |\,x_t=x\rangle_1\nonumber\\
&=& \langle e^{D_1(\omega)-D_2(\omega)  + \frac 1{2} [S_2(\omega) - S_1(\omega)]}\, |\,x_t=x\rangle_1
\end{eqnarray}
where we applied formula \eqref{dyen}.
The differences in \eqref{ppt} can be written out for a small change from the first to the second process, taking the linear order in the derivatives of $D$ and $S$ with respect to coupling $\lambda$ and probe positions $Y$, 
\begin{equation}\label{eg1}
e^{-{\cal A}} =  1- (\lambda_2-\lambda_1)\left[\frac{\partial D}{\partial \lambda}  - \frac 1{2}\frac{\partial S}{\partial \lambda}\right]_{(\lambda_1,Y^{(1)})} - (Y^{(2)}- Y^{(1)})\cdot\left[\nabla_Y D - \nabla_Y\frac{S}{2}\right]_{(\lambda_1,Y^{(1)})}
\end{equation}
 When combining \eqref{eg1} with \eqref{ppt} and plugging that into \eqref{staf}  we get the $(Y,\lambda)$-dependencies in the force on the $\alpha$th probe as
\begin{eqnarray}\label{summar}
&&f_\alpha^{\lambda+\id \lambda}(Y+\id Y) - f_\alpha^{\lambda}(Y) = -\lambda\sum_i\int \id x \,\rho_Y^\lambda(x)\,u'(x(i)-Y_\alpha)\times\\
&&\left[ \left< \id \lambda
\frac{\partial D}{\partial \lambda}  +\id Y\cdot\nabla_Y D\,\big|\,
x_t=x \right> 
- \left<\frac{\id \lambda}{2}\frac{\partial S}{\partial \lambda} +\frac{\id Y}{2}\cdot\nabla_Y  S\,\big|\,
x_t=x \right> \,\,\right]\nonumber
\end{eqnarray}
 where the averages $\langle\,\cdot\,\big|\,
 x_t=x \rangle$ are over the nonequilibrium bath with trajectories during $[0,t]$ and conditioned on the future $x_t=x$, for fixed coupling $\lambda$ and probe positions $Y$.\\
 The frenetic and entropic terms in the right-hand side of \eqref{summar} are of a totally different character. It does make sense to assume that the variable entropy flux $S(\omega)$ per $k_B$ only depends on the coupling with the probe via the interaction energy,
 \begin{eqnarray}
 \frac{\partial S}{\partial \lambda} &=& \beta\sum_{i,\alpha} \left( u(x_0(i)-Y_\alpha) - u(x_t(i)-Y_\alpha)\right)\nonumber\\
\frac{\partial S}{\partial Y_\alpha} &=& \beta\lambda\sum_{i} \left(  u'(x_t(i)-Y_\alpha) -  u'(x_0(i)-Y_\alpha)\right)\label{seas}
 \end{eqnarray}
 That means to suppose that the presence of the (static) probes does not change the nonequilibrium driving.  The position of the probes only matters for the changes in energy (purely thermodynamic).  For example, the applied nonconservative forces or the nature of the variable current does not change with $(Y,\lambda)$. On the other hand, for the first path-average in the right-hand side of \eqref{summar}, the frenesy as function on trajectories could drastically change because of the presence of probes; not only do energies get shifted but also the escape rates during a trajectory may change by the positions of the probes. Clearly, literally, probes may block bath particles.  One can think of them as static disorder possibly responsible for trapping of the bath particles.  Moreover, in both averages of \eqref{summar} there is a conditioning on the (future) event $x_t=x$, specifying where the bath-particle must be at time $t$. Again, for the entropy fluxes, as visible from \eqref{seas}, that is not a problem since we are dealing with a temporal difference of state functions.  We can essentially plug in \eqref{seas} for the entropic term in \eqref{summar}. Yet again that is not at all the case for the frenetic term as the frenesy remains path-dependent and where the conditioning on the future is essential.
 As we have seen already in Example \ref{crex} the frenetic contribution, in as far as it is ``twisted'' with respect to the entropy fluxes can be followed to be responsible for nongradient aspects in the statistical force.  We refer to \cite{nong} for the continuation of that analysis including an alternative approach using the Komatsu-Nakagawa formula of \cite{naoko}.  At any rate, the upshot of the analysis enabled by the response formula \eqref{summar} is to discuss stability around fixed points and its dependence on coupling with nonequilibria, \cite{nong,stf}.

\section{Smooth dynamical systems}\label{dynsmf}

The thermodynamic formalism relates concepts and mathematics of thermodynamics and statistical mechanics with the theory of dynamical systems.  There emerge analogues of free energy, entropy, Gibbs measures and their variational principles. That relation and mutual influence between thermodynamics and dynamical systems is old, and goes at least back to work of Helmholtz and Boltzmann, but it has been revived since the 1970-1990's.  It is not our goal to give its history or to present a systematic account; there are many other sources including \cite{dorf,katok,ruthermo,eva,galla}.  We only add the question about a possible analogue in dynamical systems for the concept of frenesy, and at the end we highlight some new developments focusing on finite time escape rates.  In contrast indeed with statistical thermodynamics, the theory of dynamical systems (e.g. in definitions of Lyapunov exponents) has concentrated mostly on taking the infinite time limit (before the macroscopic limit), which is a conceptual problem for relating it to physical systems.\\

It is fair to say that the study of steady nonequilibria has concentrated mostly on entropy production and time-irreversibility. After all those were the natural concepts arising from studies on the return to equilibrium.  In the theory of dynamical systems, it was easy enough to relate those with notions as  Kolmogorov-Sinai entropy and Policott-Ruelle resonances.  For smooth (Axiom A) dynamics one found that breaking of time-reversal invariance manifests itself statistically because of a different behavior along the stable and unstable manifolds; see e.g. \cite{g1,g2} for more details and illustrations.  The production of entropy for example has been associated to that difference. Corresponding Sinai-Ruelle-Bowen (SRB) measures, which are often called natural nonequilibrium steady states, are smooth along unstable directions and fractal in the stable direction; see also \cite{gasp,katok}.  That is thought essential by those who seek foundations of statistical mechanics in the theory of dynamical systems to circumvent the usual invariance of the Shannon entropy corresponding to a smooth phase space density evolving according to the Liouville equation.  At the same time one was looking in the Lyapunov spectrum to unravel important aspects of diffusion and transport.\\

To make the issue more specific we assume a reversible smooth dynamical system $x\mapsto \varphi(x)$ on a compact and connected manifold $\cal M$. The $\varphi$ is a diffeomorphism of $\cal M$ and reversibility means here the existence of another involutive diffeomorphism $\pi, \pi^2=1$ with $\pi\varphi\pi =\varphi^{-1}$.  
Chaoticity is translated in the requirement of having uniform hyperbolicity (transitive Anosov system). There is then a unique SRB-state with expectations
 \begin{equation}\label{srbs}
 \langle G \rangle = \lim_{\tau\rightarrow\infty} \frac 1{\tau}\,\sum_{t=0}^\tau G(\varphi_tx)
 \end{equation}
which realizes its expectations $\langle \cdot\rangle$ as time-averages for almost every randomly chosen initial point $x\in \cal M$. We took here discrete time $t$ and $\varphi_tx$ is the point at time $t$ after $t$-fold composition of $\varphi$. When phase space is contracting under these changes of variables, we speak about a dissipative dynamics.   To have dissipation one must prove (or assume) that the contraction rate is strictly negative.\\ 
 For Anosov diffeomorphisms $\varphi$ the SRB distribution is a Gibbs
 measure for the potential
  \[
 U(x) = -\log ||{\text D}\varphi_{|E^u(x)}||
 \] 
where $E^u(x)$ is the unstable subspace of the tangent space at the point $x$. 
The antisymmetric part of that potential under time-reversal gives (minus) the phase space contraction rate $\sigma(x) = -\log ||{\text D}\varphi(x)||$ in state $x$, which is formally identified with entropy production rate, \cite{GC,rue}. Time-averages of that entropy production rate as in \eqref{srbs} give the mean or stationary entropy production rate, and fluctuation theorems study symmetries in their fluctuations.  The stationary ``entropy production'' is also given by ``the sum of the positive Lyapunov exponents for the forward minus the sum of the positive Lyapunov exponents for the time-reversed dynamics,'' which is
 \[
 \sum_i \lambda^+_\varphi(i) - \sum_i \lambda^+_{\varphi^{-1}}(i) =  \sum_i [\lambda^+_\varphi(i) + \sum_i \lambda^-_\varphi(i)]
 \]
 the sum of (all) the Lyapunov exponents; see \cite{rue,verbit,stat}.\\

The question we pose here is whether the above thermodynamic analogue (formalism) for dynamical systems has a kinetic extension to include the frenetic contribution.  The frenesy in the stationary condition is then supposed to correspond likewise, to
 \[
  \frac{\langle D\rangle}{t} \rightarrow\sum_i \lambda^+_\varphi(i) + \sum_i \lambda^+_{\varphi^{-1}}(i) =  \sum_i [\lambda^+_\varphi(i) - \sum_i \lambda^-_\varphi(i)]
  \]
  the sum of positive minus the sum of negative Lyapunov exponents. That will of course change under perturbations of the dynamics $\varphi$, giving rise to excess frenesy as usual.  We are not aware of explicit studies, e.g. in linear response theory around SRB-measures which has been investigated in much mathematical detail \cite{srlin,viv}, that identify or confirm that suggested frenetic contribution in physical examples.\\

  Let us finally refer to a collection of work over the last decade in open dynamical systems. There, a central quantity indeed is the escape rate.  One considers a ``hole,'' which is a positive measure subset of the phase space.  For ergodic dynamical systems, trajectories eventually leak out and escape rates tell us how the survival probability decays asymptotically in time. What is new is to ask how the finiteness and the position of the hole may matter, and how survival and hitting probabilities behave for finite times. In \cite{buni1} it was shown that for very chaotic systems  (the most uniformly hyperbolic dynamic)  escape rates are indeed generally different for different holes and
relations between corresponding survival probabilities can be established for all
  moments of time. We refer to \cite{buni2} for the first rigorous results in the mathematical
  theory of finite time dynamics of (strongly) chaotic systems.

\section{Experimental observations}\label{expf}

There are few experimental papers up to now that explicitly look at the frenetic contribution in nonequilibrium set-ups.  Yet, with today's possibilities of trajectory visualization, image analysis and data selection, and with many refined technologies (e.g. via optical tweezers) to manipulate probes on the mesoscopic level, the experimental challenge of measuring frenesy is there to be met.\\

  A first paper is \cite{juan} where a driven Brownian particle in a
toroidal optical trap is studied for its linear response of the potential energy.  It is proven there that the frenetic contribution to the response is independently (or separately) measurable.  It was the first paper to show the experimental feasibility of the entropic--frenetic dichotomy in the study of the linear response
of nonequilibrium micron-sized systems with a small number of degrees of freedom
immersed in simple fluids.\\

The paper \cite{gal} uses the theory of linear response around nonequilibria to probe active forces in
living cells.  That is an inverse use of the response theory in Section \ref{res}, where by measuring the force, one obtains the correlation between force and displacement which is exactly the frenetic part in \eqref{genresult5}.\\

The challenge of measuring the frenetic contribution in second order response around equilibrium was taken in \cite{urna16}.  It is concerned with a colloidal particle in an anharmonic potential.  The trajectory of the particle is measured using the method of total internal
reflection microscopy.  The perturbation is a light force on the particle.\\

Finally, in \cite{urna18} the typical problem is addressed of getting ``enough'' kinetic information in many-body systems to evaluate e.g. the frenetic contribution.  The context is again that of nonlinear response theory around equilibrium but now applied to many--body systems. Issues of coarse-graining and hidden variables are obviously not only relevant for comparing theory with experimental results; such problems take place in simulation and numerical studies as well.

\section{Conclusions and outlook}

As the possibilities of observation and manipulation of mesoscopic kinetics are growing sensationally, new demands are made for the theory to summarize more systematically the relevant dynamical ensembles.  Those ensembles are specified by an action on path-space, giving the weight of the various possible paths, i.e., of the trajectories of the considered dynamical variables.  The most immediate and relevant decomposition of the action appears to be in antisymmetric and symmetric parts with respect to the time-reversal transformation on system trajectories.  Under local detailed balance the antisymmetric part in the action relates to the entropy flux (per $k_B$) into the environment, and has been studied mostly from the point of view of thermodynamics on path-space. Entropy production provides then a precise measure of time-reversal breaking.  The time-symmetric part however complements that analysis in essential ways, especially outside the close-to-equilibrium regime.  It is therefore useful to name it and to see its role and appearance in response and fluctuation theory.  This review on {\it frenesy} has sketched the main instances of that analysis mostly in the context of {\it ideal} systems, neglecting interesting issues of collective behavior and phase transitions.\\

We conclude that progress in nonequilibrium physics will also depend on the identification of time-symmetric parameters and time-symmetric (path-)observables.  Their role and nature are essential and in fact connect with older ambitions of kinetic theory. For biological processes for example, frenesy may give then a new meaning to the out-dated concept of {\it vis viva}, as the dynamical time-symmetric activity in open systems. The concept of effective temperature may be seen as a multiplicative abbreviation in an effective Kubo formula for incorporating the (additive) frenetic contribution in linear response theory.\\
The previous two sections already contained an outlook to the theory of dynamical systems and to experimental validation.  Future experiments will decide on the operational value of frenesy, complementing the 19$^\text{th}$ century concept of entropy and its relation with heat and work.    That is certainly a challenge for the study of quantum systems as well. There also however much can be hoped from a trajectory-based physics to formulate fluctuation and response relations, and to understand quantum frenetic aspects.  After all, we would speculate that the quantum influence on nonequilibrium phenomena may be most pronounced in the time-symmetric fluctuation sector.  As a final outlook and question, still for weakly interacting components, we wonder about the role of dynamical activity in learning and pattern recognition.  By relying on gradient flow in free energy landscapes, the largest part of neuro-computational models do not typically use realistic biologically-plausible processes, while time ``in our brain'' is probably again nothing but movement; see e.g. \cite{ama} for a recent and workable encoding of time in a neural network.


\newpage

\begin{thebibliography}{apa}
	
\bibitem{dGM}
S.R.~de Groot and P.~Mazur, {\it Non-equilibrium thermodynamics}, North-Holland, Amsterdam, 1962.

\bibitem{rol}
\'E.~Rold\'an, J.~Barral, P.~Martin, J.~M.R.~Parrondo and F.~J\"ulicher, Arrow of Time in Active Fluctuations.  arXiv:1803.04743v3 [cond-mat.stat-mech].

\bibitem{bor}
M.~B\"orsch, P.~Turina, C.~Eggeling, J.R.~Fries, C.A.M.~Seidel, A.~Labahn and P.~Gr\"aber,
Conformational changes of the H$^+$-ATPase from Escherichia coli upon nucleotide binding detected by single molecule fluorescence.
FEBS Letters {\bf 437}, 251--254 (1998).

\bibitem{mud}
H.S.~Muddana, S.~Sengupta, T.~E.~Mallouk, A.~Sen and P.J.~Butler,
Substrate Catalysis Enhances Single-Enzyme Diffusion.
J. Am. Chem. Soc. {\bf 132}, 2110--2111 (2010).

\bibitem{poincare}
C.~Maes, On the origin and the use of fluctuation relations for the entropy. S\'eminaire Poincar\'e {\bf 2}, 29--62 (2003). 
 
\bibitem{time}
C. Maes and  K.~Neto\v{c}n\'y, Time-reversal and Entropy.
 J. Stat. Phys. {\bf 110}, 269 (2003).

\bibitem{gibbs}
C.~Maes, The fluctuation theorem as a Gibbs property. J. Stat. Phys. {\bf 95}, 367--392 (1999).

\bibitem{crooks}
G.E.~Crooks, Nonequilibrium measurements of free energy differences for microscopically
reversible Markovian systems. J. Stat. Phys. {\bf 90}, 1481 (1998).

\bibitem{jmp2000}
C. Maes, F. Redig and A. Van Moffaert, On the definition of entropy production via examples.
J. Math. Phys. {\bf 41}, 1528--1554 (2000).

\bibitem{hthms}
W.~De Roeck, C.~Maes and K.~Neto\v{c}n\'y, H-Theorems from Macroscopic Autonomous Equations. J. Stat. Phys. {\bf 123}, 571--584 (2006).

\bibitem{3lev2}
M.~Fannes, C.~Maes and A.~Verbeure, Eds. {\it On Three Levels.
Micro-, Meso-, and Macro-Approaches in Physics}.  Springer, 1994.

\bibitem{ott}
H.C.~\"Ottinger, {\it Beyond Equilibrium Thermodynamics}. Wiley, New York (2005).

\bibitem{gen1}
R.~Kraaij, A.~Lazarescu, C.~Maes and M.A.~Peletier, Deriving GENERIC from a generalized fluctuation symmetry. 
Journal of Statistical Physics 170, 492--508 (2018).

\bibitem{sst}
S.-i.~Sasa and H.~Tasaki, Steady State Thermodynamics.  J. Stat. Phys. {\bf 125}, 125--224 (2006).

\bibitem{stefan}
C.~Maes and S.~Steffenoni, Friction and noise for a probe in a nonequilibrium fluid.  Phys. Rev. E {\bf 91}, 022128-7 (2015).

\bibitem{ext}
D.~Ruelle, Conversations on Nonequilibrium Physics with an Extraterrestrial. Physics Today {\bf 57}(5), 48 (2004).

\bibitem{go}
P.G.~Bergmann and J.L.~Lebowitz, New Approach to Nonequilibrium Process. Physical Review {\bf 99}, 578--587 (1955).

\bibitem{hal}
H.~Tasaki, Two theorems that relate discrete stochastic processes to microscopic mechanics. 
arXiv:0706.1032v1 [cond-mat.stat-mech].

\bibitem{derrida}
B.~Derrida, Non-equilibrium steady states:fluctuations and large deviations of the density and of the current. J. Stat. Mech. P07023 (2007).

\bibitem{leb}
S.~Katz, J.L.~Lebowitz, and H.~Spohn, Stationary nonequilibrium states for stochastic
lattice gas models of ionic superconductors. J. Stat. Phys. {\bf 34}, 497-–537 (1984). ---, Phase Transitions in Stationary Non-equilibrium States of Model lattice Systems. Physical Review B {\bf 28}, 1655--1658 (1983).

\bibitem{snak}
J.~Schnakenberg, Network theory of behavior of master equation systems. Rev. Mod.
Phys. {\bf 48}, 571--585 (1976).

\bibitem{Fer}
E.~Fermi, On the Origin of the Cosmic Radiation. Physical Review {\bf 75}, 1169--1174 (1949).

\bibitem{hopfield}
J.~J.~Hopfield, Kinetic Proofreading: A New Mechanism for Reducing Errors in Biosynthetic Processes Requiring High Specificity.
Proc. Nat. Acad. Sci. USA
{\bf 71}, 4135--4139 (1974).

\bibitem{land1975}
R.~Landauer, Inadequacy of entropy and entropy derivatives in characterizing
the steady state. Phys. Rev. A. {\bf 12}, 636--638 (1975).

\bibitem{heatbounds}
C.~Maes and K.~Neto\v{c}n\'y, Heat bounds and the blowtorch theorem. Annales Henri Poincar\'e {\bf 4}, 1193--1202 (2013).

\bibitem{urna} 
U.~Basu and C.~Maes, Nonequilibrium Response and Frenesy. J. Phys.: Conf. Ser. {\bf 638}, 012001 (2015).

\bibitem{winny}
C.~Maes, K.~Neto\v{c}n\'y and W. O'Kelly de Galway, Low temperature behavior of nonequilibrium multilevel systems.
Journal of Physics A: Math. Theor. {\bf 47}, 035002 (2014).

\bibitem{hang}
P.~H\"anggi, P.~Talkner and M.~Borkovec, Reaction-rate theory: fifty years after Kramers.
Rev. Mod. Phys. {\bf 62}, 251 (1990).

\bibitem{nonMax}
T.~Demaerel, W.~De Roeck and C.~Maes, Producing suprathermal tails in the stationary velocity distribution.  arXiv:1903.02312v1 [cond-mat.stat-mech]. Accepted for publication in Physica A (2019).  doi.org/10.1016/j.physa.2019.122179.

\bibitem{peir}
V.~Pierrard and M.~Lazar, Kappa Distributions: Theory and Applications in Space
Plasmas. Solar Phys. {\bf 267}, 153--174 (2010).

\bibitem{nong} 
C. Maes and K.~Neto\v{c}n\'y, Nonequilibrium corrections to gradient flow.  Chaos {\bf 29}, 073109 (2019).

\bibitem{kel}
C.~Maes and W.~O'Kelly de Galway, On the kinetics that moves Myosin V. Physica A: Statistical Mechanics and its Applications {\bf 436}, 678--685 (2015).

\bibitem{zia} 
R.K.P.~Zia, E.L.~Pr{\ae}stgaard and O.G.~ Mouritsen, Getting more from pushing less: Negative specific heat and conductivity in nonequilibrium steady states.  Am. J. Phys. {\bf 70}, 384 (2002).

\bibitem{springer}
C.~Maes, Non-Dissipative Effects in Nonequilibrium Systems. SpringerBriefs in Complexity,  ISBN 978-3-319-67780-4 (2018).

\bibitem{pri}
The more complete citation is ``{\it When all else is asleep, Time is awake, Time is irresistible.}'' Stri Parva  --- The Mahabharata, Translated by Kisari Mohan Ganguli (1889), Ch. 2, p6.

\bibitem{fac} 
P.~Faccioli, M.~Sega, F.~Pederiva and H.~Orland, Dominant Pathways
in Protein Folding. Phys. Rev. Lett. {\bf 97}, 108101 (2006).

\bibitem{cha}
L.O.~Hedges, R.L.~Jack, J.P.~Garrahan and D.~Chandler, Dynamic Order-Disorder in Atomistic Models of Structural Glass Formers. Science {\bf 323}, 1309--1313 (2009).

\bibitem{girs}
I.V.~Girsanov, On transforming a certain class of stochastic processes by absolutely continuous substitution of measures. Theory Probab. Appl. {\bf 5}, 285--301 (1960).

\bibitem{heatcond}
C.~Maes and K.~Neto\v{c}n\'y and M. Verschuere, Heat Conduction Networks. J. Stat. Phys. {\bf 111}, 1219--1244 (2003).

\bibitem{hade}
K.P.~Hadeler, {\it Topics in Mathematical Biology}.  Lecture Notes on Mathematical Modelling in the Life Sciences, Springer, 2018.

\bibitem{mer}
M.~Merolle, J.P.~Garrahan and D.~Chandler, Space–time thermodynamics of the glass transition.
PNAS {\bf 102}, 10837--10840 (2005).	

\bibitem{viv}
V.~Lecomte, C.~ Appert-Rolland and F.~van Wijland, Thermodynamic Formalism for Systems with Markov Dynamics. J. Stat. Phys. {\bf 127}, 51--106  (2007).

\bibitem{gar2}
J.P.~Garrahan, R.L.~Jack, V.~Lecomte, E.~Pitard, K.~van Duijvendijk and F.~van Wijland, Dynamical First-Order Phase Transition in Kinetically Constrained Models of Glasses.
Phys. Rev. Lett. {\bf 98}, 195702 (2007). 

\bibitem{gar3}
J.P.~Garrahan, R.L.~Jack, V.~Lecomte, E.~Pitard, K.~van Duijvendijk and F.~van Wijland,  First-order dynamical phase transition in models of glasses: an approach based on ensembles of histories. J.Phys. A: Math. Theor. {\bf 42}, 075007 (2009).

\bibitem{maarten}
C.~Maes and M.H.~van Wieren, Time-symmetric fluctuations in nonequilibrium systems.
Phys. Rev. Lett. {\bf 96}, 240601 (2006).

\bibitem{fdr}
M.~Baiesi, C.~Maes and B.~Wynants, Fluctuations and response of nonequilibrium states. Phys. Rev. Lett. {\bf 103}, 010602 (2009).

\bibitem{wojciech}
W.~De Roeck and C.~Maes, Symmetries of the ratchet current. Phys. Rev. E {\bf 76}, 051117 (2007).

\bibitem{carlo}
J.~Hooyberghs, C.~Vanderzande, Thermodynamics of histories for the one-dimensional contact process. J. Stat. Mech. P02017 (2010).

\bibitem{jar}
C.~Jarzynski, Comparison of far-from-equilibrium work relations, Comptes Rendus Physique {\bf 8}, 495 (2007).

\bibitem{kroy} 
G.~Sauermann, K.~Kroy and H.J.~Herrmann, Continuum saltation model for sand dunes.  Phys. Rev. E{\bf 64}, 031305 (2001).

\bibitem{rold}
\'E.~Rold\'an and P.~Vivo, Exact Distributions of Currents and Frenesy for Markov Bridges.  arXiv:1903.08271v2 [cond-mat.stat-mech].

\bibitem{com}
C.~Maes and K.~Neto\v{c}n\'y, Static and Dynamical Nonequilibrium Fluctuations.
 Comptes Rendus---Physique {\bf 8}, 591 (2007).

\bibitem{revjona}
L.~Bertini, A.~De Sole, D.~Gabrielli, G.~Jona-Lasinio, C.~Landim, Macroscopic fluctuation theory.
 Rev. Mod. Phys. {\bf 87}, 593 (2015).

\bibitem{jona}
L.~Bertini, A.~De Sole, D.~Gabrielli, G.~Jona-Lasinio and C.~Landim,  Macroscopic fluctuation theory for
stationary non-equilibrium states. 
J. Stat. Phys. {\bf 107}, 635 (2002).

\bibitem{O}
L. Onsager, Reciprocal Relations in Irreversible Processes.
Phys. Rev. {\bf 87}, 405 (1931); --- {\bf 38}, 2265 (1931).

\bibitem{OM}
L. Onsager and S. Machlup, Fluctuations and Irreversible Processes.
Phys. Rev. {\bf 91}, 1505 (1953).

\bibitem{spohn}
H.~Spohn, Large scale dynamics of interacting particles. Springer-Verlag, 1991.

\bibitem{DV}
M.D.~Donsker, S.R.~Varadhan,  Asymptotic evaluation of certain Markov process
expectations for large time I. 
Comm. Pure Appl. Math. {\bf 28}, 1 (1975).

\bibitem{fen}
Jin Feng and T.G.~Kurtz, {\it Large Deviations for Stochastic Processes}, Mathematical Surveys and Monographs --- American Mathematical Society (October 31, 2006).

\bibitem{lag1}
C.~Maes,  K.~Neto\v{c}n\'y and B. Wynants, On and beyond entropy production; the case of Markov jump processes.
 Markov Processes and Related Fields {\bf 14}, 445 (2008).

\bibitem{lag2}
C.~Maes,  K.~Neto\v{c}n\'y and B. Wynants, Steady state statistics of driven diffusions.
 Physica A {\bf 387}, 2675 (2008).

\bibitem{epl}
C.~Maes and K.~Neto\v{c}n\'y, The canonical structure of dynamical fluctuations in mesoscopic nonequilibrium steady states.
Europhys. Lett. {\bf 82}, 30003 (2008).

\bibitem{MPR13}
A.~Mielke, M.A.~Peletier and D.R.M.~Renger, On the relation between gradient flows and the large-deviation principle, with applications to Markov chains and diffusion. 
{\tt arXiv:1312.7591 [math.FA]}.

\bibitem{pons}
A.~Mielke, M.A.~Peletier and D.R.M.~Renger, A generalization of Onsager's reciprocity relations to gradient flows with nonlinear mobility. 
Proceedings of IWNET 2015. {\tt arXiv:1510.06219}.

\bibitem{ncwh}
C.~Maes, F.~Redig and M.~Verschuere, No current without heat. 
J. Stat.Phys. {\bf 106}, 569 (2002).

\bibitem{fren}
C.~Maes, Frenetic Bounds on the Entropy Production. Physical Review Letters 119, 160601 (2017).

\bibitem{mono}
C.~Maes, K.~Neto\v{c}n\'y and B.~Wynants, Monotonicity of the dynamical activity. Journal of Physics A: Mathematical and General {\bf 45}, 455001 (2012).

\bibitem{minep}
C.~Maes and K.~Neto\v{c}n\'y, Minimum entropy production principle, Scholarpedia, 8(7):9664 (2013). ---, C.~Maes and K.~Neto\v{c}n\'y, Minimum entropy production principle from a dynamical fluctuation law. Journal of Mathematical Physics {\bf 48}, 053306 (2007).

\bibitem{info}
D.~Felice, C.~Cafaro and S.~Mancini, Information geometric methods for complexity. Chaos {\bf 28}, 032101 (2018).

\bibitem{terl}
I.~Di Terlizzi and M.~Baiesi, Kinetic uncertainty relation. J. Phys. A: Math. Theor. {\bf 52},  02LT03 (2019). 

\bibitem{des}
A.~Dechant and S.-I.~Sasa, Fluctuation-response inequality out of equilibrium. arXiv:1804.08250 (2018).

\bibitem{udo2} 
P.~Pietzonka, A.C.~Barato, and U.~Seifert, Universal bounds on current fluctuations.
 Phys. Rev. E 9{\bf 3}, 052145 (2016).

\bibitem{hal2}
Naoto Shiraishi, Keiji Saito and Hal Tasaki, Universal trade-off relation between power and efficiency for heat engines.
 Phys. Rev. Lett. {\bf 117}, 190601 (2016).

\bibitem{gar2}
J.P. Garrahan, Simple bounds on fluctuations and uncertainty relations for first-passage times of counting observables.
Phys. Rev. E {\bf 95}, 032134 (2017). 

\bibitem{vu}
Tan Van Vu and Yoshihiko Hasegawa, Uncertainty Relations for Underdamped Langevin Dynamics.  arXiv:1901.05715v2 [cond-mat.stat-mech].

\bibitem{green}
M.S.~Green, Markoff Random Processes and the Statistical Mechanics of Time-Dependent Phenomena. II. Irreversible Processes in Fluids. J. Chem. Phys {\bf 22}, 398 (1954); ---, Brownian Motion in a Gas of Noninteracting Molecules. J. Chem. Phys. {\bf 19}, 1036 (1951); ---, Comment on a Paper of Mori on Time-Correlation Expressions for Transport Properties. Phys.
Rev. {\bf 119}, 829 (1960).

\bibitem{kub}
R.~Kubo, Statistical-Mechanical Theory of Irreversible Processes. I. General Theory and Simple Applications to Magnetic and Conduction Problems. J. Phys. Soc. Jpn. {\bf 12}, 570--586 (1957).

\bibitem{njp}
M.~Baiesi and C.~Maes, An update on nonequilibrium linear response.
 New J. Phys. {\bf 15}, 013004 (2013).

\bibitem{urna16}
L.~Helden, U.~Basu, M.~Kr\"uger and C.~Bechinger, Measurement of second-order response without perturbation. EPL {\bf 116}, 60003 (2016).

\bibitem{juan}
J.~R.~Gomez-Solano, A.~Petrosyan, S.~Ciliberto and C.~Maes, Fluctuations and response in a non-equilibrium micron-sized system. Journal of Statistical Mechanics, P01008 (2011).

\bibitem{falb}
G.~Falasco and M.~Baiesi, Nonequilibrium temperature response for stochastic overdamped systems, New J. Phys. {\bf 18}, 043039 (2016).

\bibitem{res1} 
P.~H\"anggi, Stochastic Processes II: Response Theory and Fluctuation Theorems.  Helv. Phys. Acta 51 202--219 (1978).

\bibitem{res2}
M.~Falcioni, S.~Isola and A.~Vulpiani, Correlation functions and relaxation properties in chaotic dynamics and statistical mechanics. Phys. Lett. A {\bf 144}, 341 (1990).

\bibitem{res3}
L.~Cugliandolo, J.~Kurchan and G.~Parisi,  Off-equilibrium dynamics and aging in unfrustrated systems. J. Phys. I {\bf 4}, 1641 (1994).

\bibitem{res4}
D.~Ruelle, General linear response formula in statistical mechanics, and the fluctuation--dissipation theorem far from equilibrium. Phys. Lett. A {\bf 245}, 220--224 (1998).

\bibitem{res5}
T.~Nakamura and S.~Sasa, A fluctuation-response relation of many Brownian particles under non-equilibrium conditions. Phys. Rev. E {\bf 77}, 021108 (2008).

\bibitem{res6}
R.~Chetrite, G.~Falkovich and K.~Gawedzki, Fluctuation relations in simple examples of non-equilibrium steady states.  J. Stat. Mech. P08005 (2008).

\bibitem{res7}
T.~Speck and U.~Seifert, Restoring a fluctuation-dissipation theorem in a nonequilibrium steady state. Europhys. Lett. {\bf 74}, 391--396 (2006).

\bibitem{res8}
T.~Speck and U.~Seifert, Extended fluctuation-dissipation theorem for soft matter in stationary flow. Phys. Rev. E {\bf 79}, 040102 (2009).

\bibitem{res9}
T.~Speck and U.~Seifert, Fluctuation-dissipation theorem in nonequilibrium steady states. Europhys. Lett. {\bf 89}, 10007 (2010).

\bibitem{res10}
J.~Prost, J.F.~Joanny and J.M.~Parrondo, Generalized Fluctuation-Dissipation Theorem for Steady-State Systems. Phys. Rev. Lett. {\bf 103}, 090601 (2009).

\bibitem{res11}
G.~Verley, R.~Ch\'etrite and D.~Lacoste, Modified fluctuation-dissipation theorem near non-equilibrium states and applications to the Glauber-Ising chain. J. Stat. Mech. P10025 (2011).

\bibitem{res12}
E.~Lippiello, F.~Corberi and M.~Zannetti, Off-equilibrium generalization of the fluctuation dissipation theorem for Ising spins and measurement of the linear response function. Phys. Rev. E {\bf 71}, 036104 (2005).

\bibitem{res13}
E.~Lippiello, F.~Corberi and M.~Zannetti, Fluctuation dissipation relations far from equilibrium. J. Stat. Mech. P07002 (2007).

\bibitem{naoko}
T.S.~Komatsu and N.~Nakagawa,
An expression for stationary distribution in nonequilibrium steady state.
Phys. Rev. Lett. {\bf 100}, 030601 (2008).

\bibitem{hasa}
T.~Harada and S.-I.~Sasa,  Equality connecting energy dissipation with a violation of the fluctuation-response relation. Phys. Rev.Lett. {\bf 95}, 130602 (2005).

\bibitem{dech}
A.~Dechant and S.-I.~Sasa, Fluctuation-response inequality out of equilibrium. arXiv:1804.08250v2 [cond-mat.stat-mech].

\bibitem{cug}
L.F.~Cugliandolo, J.~Kurchan and L.~Peliti. Energy flow, partial equilibration, and effective temperatures in systems with slow dynamics. Phys. Rev. E {\bf 55}, 3898 (1994).

\bibitem{pccp}
U.~Basu, M.~Kr\"uger, A.~Lazarescu and C.~Maes, Frenetic aspects of second order response. Physical Chemistry Chemical Physics {\bf 17}, 6653-6666 (2015).

\bibitem{cpr}
L.~Agozzino and K.~Dill,
Dynamics in dissipative systems. And, their Maximum Caliber trajectories in a
solvable model.   arXiv:1904.11426v1 [cond-mat.stat-mech].

\bibitem{vanh}
L.~Van Hove, Correlations in Space and Time and Born Approximation Scattering in Systems of Interacting Particles.
Phys. Rev. {\bf 95}, 249 (1954).

\bibitem{jsp}
C.~Maes, On the Second Fluctuation--Dissipation Theorem for Nonequilibrium Baths. J. Stat. Phys. {\bf 154}, 705--722 (2014).

\bibitem{leipzig}
C.~Maes and T.~Thiery, The induced motion of a probe coupled to a bath with random resettings.  J. Phys. A: Math. Theor. {\bf 50}, 415001 (2017).

\bibitem{krueger}
M.~Kr\"uger and C.~Maes, The modified Langevin description for probes in a nonlinear medium.  Journal of Physics: Condensed Matter {\bf 29}, 064004 (2017).

\bibitem{hel}
E.~Helfand, Transport Coefficients from Dissipation in a Canonical Ensemble. Phys. Rev. {\bf 119}, 1 (1960).

\bibitem{hol}
C.S.~Holling, Resilience and stability of ecological systems. Annual Review of Ecology and Systematics {\bf 4}, 1--23 (1973). 

\bibitem{gasp}
P.~Gaspard, Chaos, Scattering and Statistical Mechanics. Cambridge University Press (Cambridge)
1998.

\bibitem{proc}
M.~Baiesi, C.~Maes and B.~Wynants, The modified Sutherland-Einstein relation for diffusive non-equilibria. Proceedings of the Royal Society A {\bf 467}, 2792--2809 (2011).

\bibitem{soghra}
C.~Maes, S.~Safaverdi, P.~Visco and F.~van Wijland, Fluctuation-response relations for nonequilibrium diffusions with memory. Physical Review E {\bf 87}, 022125 (2013).

\bibitem{gal}
P.~Bohec, F.~Gallet, C.~Maes, S.~Safaverdi, P.~Visco and F.~Van Wijland, Probing active forces via a fluctuation-dissipation relation: Application to living cells. Europhysics Letters 102, 50005 (2013).


\bibitem{sar}
A.~Sarracino, F.~Cecconi, A.~Puglisi and A.~Vulpiani, Nonlinear response of inertial tracers in steady laminar flows:
differential and absolute negative mobility. Phys. Rev. Lett. {\bf 117}, 174501 (2016). 

\bibitem{gar}
R.L.~Jack, D.~Kelsey, J.P.~Garrahan and D.~Chandler, Negative differential mobility of weakly driven particles in models of glass formers.
Phys. Rev. E {\bf 78}, 011506 (2008). 

\bibitem{negheatcap} 
P.~Baerts, U.~Basu, C.~Maes and S.~Safaverdi, The frenetic origin of negative differential response. Physical Review E {\bf 88}, 052109 (2013).

\bibitem{oliver} 
O.~B\'enichou, P.~Illien, G.~Oshanin, A.~Sarracino and R.~Voituriez.  Microscopic theory for negative differential mobility in crowded environments.
 Phys. Rev. Lett. {\bf 113}, 268002 (2014).

\bibitem{chemfalasco}
G.~Falasco, T.~Cossetto, E.~Penocchio and M.~Esposito,
Negative differential response in chemical reactions.  arXiv:1812.11245v1 [cond-mat.stat-mech].

\bibitem{hao}
Hao Ge, Min Qian and Hong Qian, Stochastic theory of nonequilibrium steady states. Part II: Applications in chemical biophysics.  Physics Reports {\bf 510}, 87--118  (2012).
 
\bibitem{muk} 
J.~Cividini, D.~Mukamel and H.A.~Posch, Driven tracer with absolute negative mobility. J. Phys. A: Math. Theor. {\bf 51}, 085001 (2018).
 
\bibitem{col}
 M.~Colangeli, C.~Maes and B.~Wynants, A meaningful expansion around detailed balance. J. Phys. A: Math. Theor. {\bf 44}, 095001 (2011)
  
\bibitem{stf}
U.~Basu, C.~Maes and K.~Neto\v{c}n\'y, How Statistical Forces Depend on the Thermodynamics and Kinetics of Driven Media. Physical Review Letters {\bf 114}, 250601 (2015).
 
\bibitem{dorf}
J.R.~Dorfman, {\it An Introduction to chaos in nonequilibrium statistical mechanics}. Cambridge
 University Press (Cambridge) 1999.
 
\bibitem{katok}
 A.~Katok and B.~Hasselblatt, {\it Introduction to the modern theory of dynamical systems}. In:Volume 54 of
 Encyclopedia of Mathematics and its Applications. Cambridge: Cambridge University Press, 1995
 
 \bibitem{ruthermo}
 D.~Ruelle,  {\it Thermodynamic formalism}. In: Volume 5 of Encyclopedia of Mathematics and its Applications.
 Reading, Mass: Addision-Wesley, 1978
 
\bibitem{eva}
D.J.~Evans, D.J.~Searles, S.R.~Williams, {\it Fundamentals of Classical Statistical Thermodynamics: Dissipation, Relaxation, and Fluctuation Theorems}. John Wiley \& Sons, 2016.

\bibitem{galla}
G.~Gallavotti, F.~Bonetto and G.~Gentile, {\it Aspects of Ergodic, Qualitative and Statistical Theory of Motion}.  Springer 2004.  See there on particular the Appendix by G.~Gallavotti on nonequilibrium thermodynamics.

\bibitem{g1}
G.~Gallavotti, Nonequilibrium Thermodynamics.  arXiv:1901.08821v1 [cond-mat.stat-mech].

\bibitem{g2}
G.~Gallavotti, Navier-Stokes equation: irreversibility turbulence and ensembles equivalence,  arXiv:1902.09610v2 [cond-mat.stat-mech].

\bibitem{GC}
G.~Gallavotti, and E.G.D.~Cohen, Dynamical ensembles in stationary states. J. Stat. Phys. {\bf 80},
931--970 (1995).
   
\bibitem{rue}
D.~Ruelle, Smooth Dynamics and New Theoretical Ideas in Nonequilibrium Statistical Mechanics,
J. Stat. Phys. {\bf 95}, 393--468 (1999).

\bibitem{verbit}
C.~Maes and E.~Verbitskiy, Large Deviations and a Fluctuation Symmetry for Chaotic Homeomorphisms,
Commun. Math. Phys. {\bf 233}, 137--151 (2003).
 
\bibitem{stat}
C.~Maes, From dynamical systems to statistical mechanics: the case of the fluctuation theorem. J. Phys. A {\bf 50}, 381001 (2017). 

\bibitem{srlin}
D.~Ruelle, A review of linear response theory for general differentiable dynamical systems.  Nonlinearity {\bf 22}, 855--870 (2009).

\bibitem{viv}
V.~Baladi, T.~Kuna and V.~Lucarini,
Linear and fractional response for the SRB measure of smooth hyperbolic attractors and discontinuous observables. Nonlinearity {\bf 30}, 1204--1220 (2017). 

\bibitem{buni1}
L.A.~Bunimovich and A.~Yurchenko, Where to place a hole to achieve maximal escape rate.
Isr. J. Math. {\bf 182}, 229--52 (2011).

\bibitem{buni2}
M.~Bolding and L.A.~Bunimovich, Where and When Orbits of Strongly Chaotic Systems Prefer to Go.  Nonlinearity {\bf 32}, 1731--1771 (2019).

\bibitem{urna18}
U.~Basu, L.~Helden, M.~Kr\"uger, Extrapolation to nonequilibrium from coarse grained response theory. Phys. Rev. Lett. {\bf 120}, 180604 (2018).

\bibitem{ama}
A.~Maes, M.~Barahona and  C.~Clopath, Learning spatiotemporal signals using a recurrent spiking network that discretizes time. doi: 10.1101/693861, 
bioRxiv 69386.


\end{thebibliography}
\end{document}